Article

# Sub-part-per-trillion test of the Standard Model with atomic hydrogen

Lothar Maisenbacher[1,4] ✉, Vitaly Wirthl[1], Arthur Matveev[1], Alexey Grinin[1,5], Randolf Pohl[2], Theodor W. Hänsch[1,3] & Thomas Udem[1,3]

Quantum electrodynamics (QED), the first relativistic quantum field theory, describes light–matter interactions at a fundamental level and is one of the pillars of the Standard Model (SM). Through the extraordinary precision of QED, the SM predicts the energy levels of simple systems such as the hydrogen atom with up to 13 significant digits[1], making hydrogen spectroscopy an ideal test bed. The consistency of physical constants extracted from different transitions in hydrogen using QED, such as the proton charge radius $r_p$, constitutes a test of the theory. However, values of $r_p$ from recent measurements[2-7] of atomic hydrogen are partly discrepant with each other and with a more precise value from spectroscopy of muonic hydrogen[8,9]. This prevents a test of QED at the level of experimental uncertainties. Here we present a measurement of the 2S–6P transition in atomic hydrogen with sufficient precision to distinguish between the discrepant values of $r_p$ and enable rigorous testing of QED and the SM overall. Our result $\nu_{2S-6P}$ = 730,690,248,610.79(48) kHz gives a value of $r_p$ = 0.8406(15) fm at least 2.5-fold more precise than from other atomic hydrogen determinations and in excellent agreement with the muonic value. The SM prediction of the transition frequency (730,690,248,610.79(23) kHz) is in excellent agreement with our result, testing the SM to 0.7 parts per trillion (ppt) and, specifically, bound-state QED corrections to 0.5 parts per million (ppm), their most precise test so far.

The binding energy of atomic hydrogen can be expressed as[1]

$$E_{nlJ} = chR_\infty \left( f_{nlJ}^{\text{Dirac}}\left(\alpha, \frac{m_p}{m_e}\right) + f_{nlJ}^{\text{QED}}\left(\alpha, \frac{m_p}{m_e}, \ldots\right) + \delta_{l0}\frac{C_{NS}}{n^3}r_p^2 \right), \quad (1)$$

in which $n$, $l$ and $J$ are, respectively, the principal, orbital and total electronic angular momentum quantum numbers of the energy level of interest. $f_{nlJ}^{\text{Dirac}}$ is the Dirac eigenvalue ($\propto 1/n^2$ in leading order), whereas the second term $f_{nlJ}^{\text{QED}}$ ($\propto 1/n^3$ in leading order) contains the corrections from bound-state QED, such as self-energy and vacuum polarization, including muonic and hadronic contributions[1]. Both terms depend on the fine-structure constant $\alpha$ and the electron-to-proton mass ratio $m_p/m_e$, which are known with sufficient accuracy from other experiments that do not require bound-state QED (refs. 1,10–14). The third term is the leading-order nuclear size correction for S-states ($l = 0$), accounting for the finite root-mean-square (rms) charge radius of the proton, $r_p$. The second and third terms constitute the Lamb shift and contribute about 1 ppm and 100 ppt, respectively, to the 2S–6P transition frequency (Extended Data Table 1). The unitless terms are converted to SI units (International System of Units) using the Rydberg constant $R_\infty$.

To compare measured energy levels or transition frequencies with equation (1), $r_p$ and $R_\infty$ must be known (speed of light in vacuum $c$ and Planck's constant $h$ are defined). In practice, $r_p$ and $R_\infty$ are largely determined from such measurements themselves and more than two measurements of distinct transitions are necessary to test equation (1).

A special case is the determination of $r_p$ with laser spectroscopy of muonic hydrogen[8,9,15]. In this exotic atom, the electron is replaced with a negative muon, whose larger mass increases $C_{NS}$ of equation (1) by four orders of magnitude, allowing a precise determination of $r_p$ without requiring other high-precision input. Because the nuclear size correction scales as $1/n^3$ as the other QED corrections, discrepant $r_p$ values from atomic and muonic hydrogen can indicate missing or incomplete QED terms (Methods). Notably, the value of $r_p$ from the muonic measurement was found to be significantly smaller (>5σ) than the then-established value (CODATA 2014 (ref. 16)).

This proton radius puzzle led to extensive research efforts[2-7,17]. We first addressed it with a precision measurement of the 2S–4P transition in atomic hydrogen[2], which favoured the muonic result, but could not conclusively (>5σ) rule out the previous value. Subsequent measurements in atomic hydrogen have followed[3-7] but they are partly discrepant with the muonic value and with each other, and none is precise enough to conclusively test the muonic value, as visualized in Fig. 1. Until now, this has prevented a verification of QED at the level of experimental uncertainties.

Here we report on laser spectroscopy of the 2S–6P transition in atomic hydrogen with sufficient precision to distinguish between the discrepant values of $r_p$. This precision corresponds to finding the transition frequency to one part in 15,000 of the experimental linewidth, to our knowledge unprecedented for laser spectroscopy, requiring a thorough understanding of any asymmetric distortions of the line

[1]Max-Planck-Institut für Quantenoptik, Garching, Germany. [2]Johannes Gutenberg-Universität Mainz, Mainz, Germany. [3]Ludwig-Maximilians-Universität München, Munich, Germany. [4]Present address: University of California, Berkeley, Berkeley, CA, USA. [5]Present address: Northwestern University, Evanston, IL, USA. ✉e-mail: lothar.maisenbacher@mpq.mpg.de

# Article

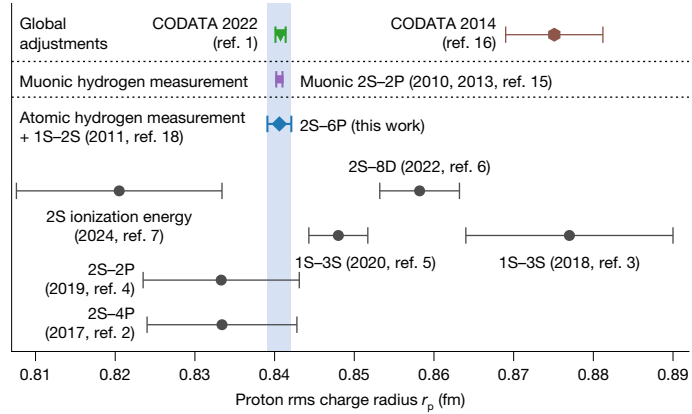

**Fig. 1 | Proton rms charge radius $r_p$.** Previous determinations of $r_p$ from atomic hydrogen spectroscopy (refs. 2–7; black circles) are partly discrepant with each other and with the value of $r_p$ from spectroscopy of muonic hydrogen (refs. 8,9,15; violet square) and therefore could not conclusively resolve the initial 5.6$\sigma$ discrepancy between the 2010 muonic value and the then-established larger value (as summarized in the CODATA 2014 global adjustment of fundamental constants[16]; brown hexagon). The value of $r_p$ from atomic hydrogen spectroscopy of the 2S–6P transition in this work (blue bar and diamond) is at least 2.5-fold more precise than other atomic hydrogen determinations and in excellent agreement with the muonic value. It disagrees with the fourfold less precise CODATA 2014 value by 5.5$\sigma$. The $r_p$ values are determined by combining each measurement with the 1S–2S transition frequency[18] and equation (1) (Pearson correlation coefficient $r$ < 0.05 between $r_p$ values). The most recent CODATA 2022 global adjustment (ref. 1; green triangle) essentially corresponds to the muonic value owing to the exceptionally low uncertainty of the latter. Error bars show one-standard-deviation uncertainties. Electron–proton scattering data are not shown, as different analyses give significantly different values of $r_p$ (refs. 1,48). Lattice QCD calculations of $r_p$ show promise but are not yet competitive[49]. See Extended Data Fig. 2 for the Rydberg constant $R_\infty$ from atomic hydrogen combined with the muonic value of $r_p$.

shape at that level and a large experimental signal-to-noise ratio. By combining our measurement with the precisely known 1S–2S transition frequency[18], we determine $r_p$ with 2.5-fold higher precision than the previous best determination from atomic hydrogen[5] (Fig. 1). Our value of $r_p$ is in excellent agreement with the muonic value but fourfold more precise than and in significant disagreement (5.5$\sigma$) with the CODATA 2014 value[16]. Consequently, we use the muonic value of $r_p$ as input to equation (1) (along with the 1S–2S transition frequency), which allows us to compare the SM prediction of the 2S–6P transition frequency with our measurement. This constitutes a test of the SM to 0.7 ppt and of bound-state QED corrections to 0.5 ppm.

## Principle of the measurement
### 2S–6P transition

We study the 2S–6P transition in a cryogenic beam of hydrogen atoms using Doppler-free one-photon laser spectroscopy. Although the transition has been previously observed with laser spectroscopy[19,20], this work presents a substantial improvement. Using a linearly polarized, 410-nm spectroscopy laser, we alternately examine two dipole-allowed transitions from the metastable initial $2S_{1/2}^{F=0}$, $m_F = 0$ level: the $2S–6P_{1/2}$ transition to the $6P_{1/2}^{F=1}$, $m_F = 0$ level and the $2S–6P_{3/2}$ transition to the $6P_{3/2}^{F=1}$, $m_F = 0$ level, as shown in Fig. 2a ($F$, total angular momentum quantum number; $m_F$, magnetic quantum number). The excited 6P levels rapidly decay, directly or through cascades, to the 1S and 2S manifolds, resulting in a $\Gamma$ = 3.90 MHz natural transition linewidth. These decays, predominantly the direct Lyman-$\varepsilon$ decay to the 1S manifold, are the experimental signal and the fluorescence line shape is observed in line scans by recording this signal at different spectroscopy laser detunings. A fraction of $\gamma_{ei}/\Gamma$ = 3.9% or 7.9% of decays from the excited level lead back to the initial 2S level for the $2S–6P_{1/2}$ or $2S–6P_{3/2}$ transitions, respectively. Notably, quantum interference (QI) between excitation–decay paths that go through either excited level but lead to the same final level can cause substantial distortions of the fluorescence line shape. The associated line shifts are on the order of $\Gamma^2/\Delta\nu_{FS}(6P) \approx \Gamma/100$ (refs. 2,21,22), in which $\Delta\nu_{FS}(6P) \approx 405$ MHz is the 6P fine-structure splitting between the excited levels. Because the magnitude and sign of the distortions depend on the detection direction (relative to the laser polarization), here we use a large detection solid angle and a magic polarization angle to strongly suppress the shift[22].

### Experimental apparatus

The key in-vacuum components of the experimental apparatus (described in detail in ref. 23) are shown in Fig. 2b. Briefly, a cryogenic beam of hydrogen atoms is formed by a copper nozzle (circular aperture with $r_1 = 1$ mm radius) held at temperature $T_N = 4.8$ K. The atoms are prepared in the initial $2S_{1/2}^{F=0}$, $m_F = 0$ level by Doppler-free two-photon excitation from the 1S ground level with a preparation laser (243 nm wavelength; Fig. 2a,b), collinear with the atomic beam. The divergence of the atomic beam is limited to approximately 10 mrad in the transverse ($x$) direction by a collimating aperture (1.2 mm width, placed 154 mm after the nozzle).

The atomic beam enters the 2S–6P spectroscopy region inside a cylindrical detector assembly, in which, at a distance $L = 204$ mm from the nozzle, it crosses counterpropagating (along $x$) spectroscopy laser beams at an adjustable atomic beam offset angle $|\alpha_0| = 0$–12 mrad from the orthogonal. In the ideal case of laser beams with identical wavefront curvature and power, this excitation scheme produces a line shape whose centre of mass is free of first-order Doppler shifts (but not necessarily free of Doppler broadening), as the interaction with the respective beams results in Doppler shifts of equal magnitude but opposite sign. Here an active fibre-based retroreflector (AFR)[24–26], consisting of polarization-maintaining fibre, collimator and high-reflectivity mirror, generates the required high-quality, wavefront-retracing beams (2.2 mm $1/e^2$ intensity radius; $P_{2S-6P}$ = 5–30 µW power in each beam). The AFR is attached to the rotatable detector cylinder, allowing an in situ adjustment of $\alpha_0$ (1 mrad accuracy). We align $\alpha_0$ close to zero to avoid splitting the line shape into two Doppler components from the counterpropagating beams, except when characterizing the light force shift (LFS; see below). The polarization angle $\theta_L$ of the laser beams, relative to the axis of the cylinder (along $y$), is set to 56.5°, the magic angle at which QI distortions are suppressed, or orthogonal to it (146.5°).

The photons emitted by the 6P decay eject photoelectrons from the cylinder walls, which are drawn to and counted with channel electron multipliers at the top and bottom of the cylinder (top and bottom detectors). Each detector covers >20% of the total solid angle, further suppressing QI distortions. A segmented Faraday cage surrounds the spectroscopy region, shielding external electric fields and allowing the application of bias fields to characterize stray electric fields and the dc-Stark shift caused by such fields (Methods).

### Doppler-free one-photon spectroscopy

We investigate different atomic velocity groups $\tau_i$ ($i$ = 1–16; Extended Data Table 2) by periodically blocking the 1S–2S preparation laser, thereby intermittently stopping the production of 2S atoms, and recording the signal as a function of delay time $\tau$ = 10–2,560 µs. The longer the $\tau$, the slower the 2S atoms contributing to the signal. The speed distribution of atoms contributing to the signal is well described by a Maxwell–Boltzmann flux distribution with an extra multiplicative factor $\exp(-v_{cut-off}/v)$ with a characteristic cut-off speed $v_{cut-off} \approx 50$ m s$^{-1}$, accounting for the loss of slower atoms through collisions (Methods). The mean speeds $\bar{v}$ of the velocity groups cover a wide range (253(5) to 65(1) m s$^{-1}$; overall mean speed $\langle\bar{v}\rangle$ = 195(6) m s$^{-1}$) and their transverse velocities are approximately Gaussian distributed with a full width at half maximum (FWHM) ranging from 3.4 to 0.6 m s$^{-1}$ (parentheses give the standard deviation over the data groups of Extended Data Table 3).

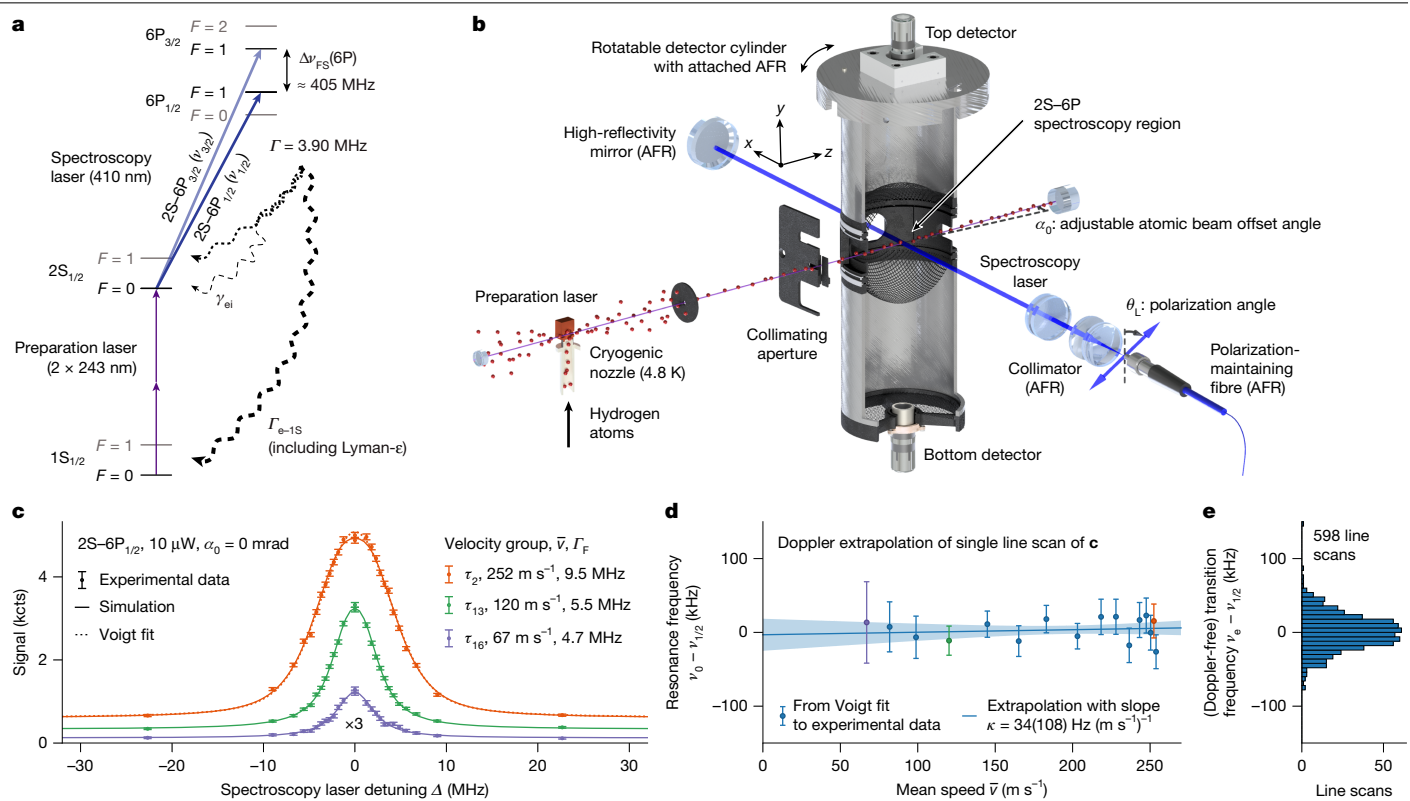

**Fig. 2 | Doppler-free one-photon spectroscopy of the 2S–6P transition. a**, Relevant level scheme of atomic hydrogen (not to scale). Solid arrows indicate laser-driven transitions and dashed arrows indicate spontaneous decay. Spectroscopy laser is on resonance with 2S–6P$_{1/2}$ (dark blue) or 2S–6P$_{3/2}$ (light blue) transitions. Levels shown in grey are not resonantly coupled by lasers. **b**, Key components of experimental apparatus (to scale, cutaway). Dashed black lines (orthogonal to spectroscopy laser) visualize atomic beam offset angle $\alpha_0$ (exaggerated) and laser polarization angle $\theta_L$. **c**, Typical line scan of the 2S–6P$_{1/2}$ transition (acquired within 40 s; full detuning range is ±50 MHz; Methods). The fluorescence signal (top detector; kcts, kilocounts) is recorded for different velocity groups $\tau_i$ with mean speeds $\bar{v}$, with three shown here (orange, green and purple circles; $\tau_{16}$ scaled (×3) for visibility). Error bars show expected one-standard-deviation ($\sigma$) shot noise. The FWHM linewidth $\Gamma_F$ reduces for slower velocity groups as Doppler broadening reduces. Solid lines show simulated line shape (scaled and offset to match the signal) and the dotted orange line shows the Voigt line shape fit to $\tau_2$ (see Extended Data Fig. 1 for fit residuals). $P_{2S–6P} = 10$ μW spectroscopy laser power and $\alpha_0 = 0$ mrad were used. **d**, Resonance frequencies $\nu_0$ (circles) determined from Voigt line shape fits to velocity groups of scan of panel **c** versus $\bar{v}$. Error bars show 1$\sigma$ fit uncertainty. Extrapolation to zero speed (blue line; blue shading, 1$\sigma$ confidence interval) gives Doppler-free transition frequency $\nu_e$ and Doppler slope $\kappa$. $\nu_0$ has been corrected for light force, QI and second-order Doppler shifts and all other corrections. **e**, Histogram of all 598 detector-averaged Doppler-free transition frequencies $\nu_e$ (data group G3 of Extended Data Table 3) determined with the same experimental parameters as the line scan of panels **c** and **d**.

Figure 2c shows the top detector signal for three velocity groups (circles) for a typical line scan of the 2S–6P$_{1/2}$ transition. Fast velocity groups, such as $\tau_2$ (orange circles), have a FWHM linewidth $\Gamma_F$ substantially larger than the natural linewidth $\Gamma$, owing to Doppler broadening from their large transverse velocity widths. Conversely, for slower velocity groups (green and purple circles), $\Gamma_F$ approaches $\Gamma$ as the transverse velocity width decreases. Overall, $\Gamma_F$ ranges from 9.6(4) to 5.5(6) MHz (Extended Data Table 2).

Our line shape simulations (solid lines; here the QI model is shown; Methods) are in excellent agreement with the experimental data, reproducing the Doppler broadening with $v_{cut\text{-}off}$ as the only free parameter. The resonance frequency $\nu_0$ of each velocity group is determined by fitting simple Voigt or Voigt doublet line shapes (dotted line; Methods) to the data. This differs from the approach taken in our previous 2S–4P measurement[2], in which a theoretically motivated, asymmetric line shape model was used to account for QI distortions. Here the QI distortions are much smaller, owing to the magic polarization angle and large detection solid angle, and distortions from the LFS dominate (Extended Data Fig. 1), for which no equivalent line shape model is known to us. Instead, we fit the same simple line shape model to the simulations and correct the experimental resonance frequency with our LFS and QI simulations.

To remove any residual first-order Doppler shift, we use the linear model $\nu_0 = \nu_e + \kappa \bar{v}$ to extrapolate the resonance frequencies $\nu_0$ of each line scan to zero speed (Fig. 2d). This gives the Doppler-free transition frequency $\nu_e$, the Doppler slope $\kappa$ and the resulting effective frequency correction $\Delta\nu_e = -\kappa \langle \bar{v} \rangle$ (Methods). Figure 2e shows $\nu_e$ of all 598 line scans recorded with the same experimental parameters as in Fig. 2c,d. In total, the dataset presented here contains 3,155 line scans, acquired in three measurement runs (A, B and C) and grouped into 17 experimental parameter combinations (data groups; Extended Data Table 3).

## Light force shift

Just as light waves diffract on a periodic structure, matter waves can diffract on the periodic structure formed by a standing wave of light, as first predicted for electrons by Kapitza and Dirac[27]. Here a similar diffraction can occur as the atoms, which may be understood as matter waves, cross the standing intensity wave formed by the counter-propagating spectroscopy laser beams used to suppress the first-order Doppler shift (Fig. 2b). This effect, along with scattering on the standing wave, leads to a distortion of the fluorescence line shape and a line shift (LFS). Although such shifts have been observed in other experiments using standing waves[28–30], the behaviour and size of the shift and the necessary theoretical treatment are highly dependent on the exact experimental conditions. A travelling wave can be used to avoid the LFS[31] but this approach has not yet demonstrated the level of Doppler shift suppression required here.

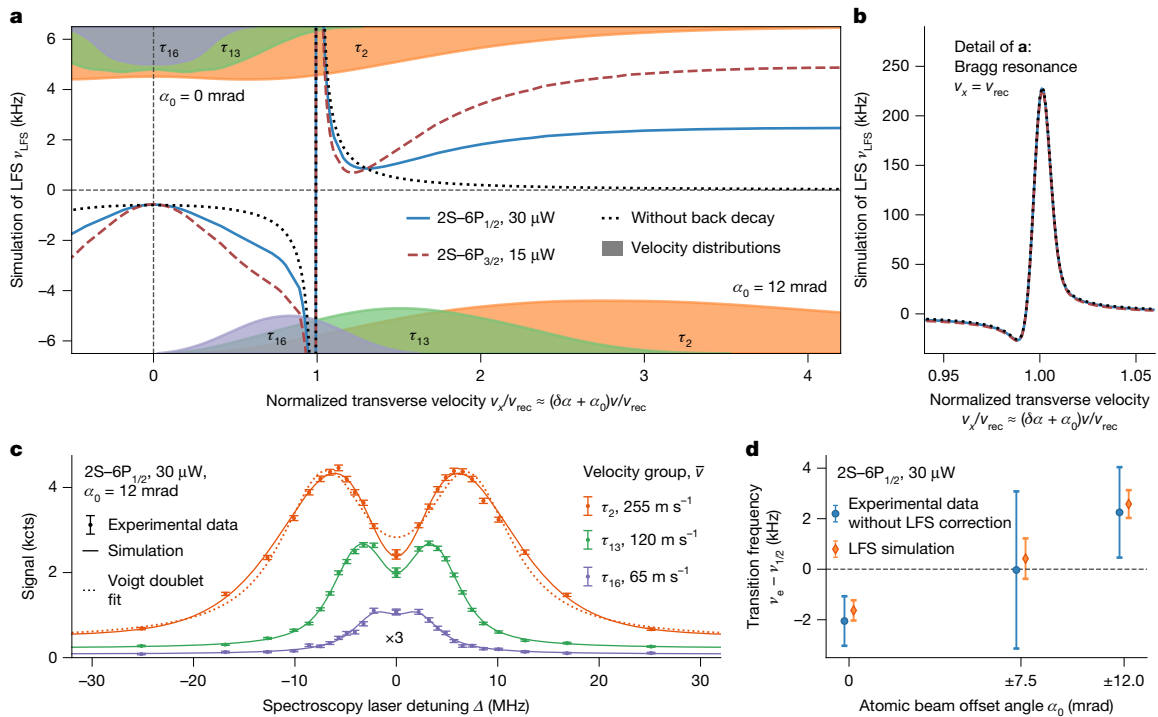

**Fig. 3 | Simulation and measurement of LFS. a**, Simulation of LFS $\nu_{LFS}$, resulting from interaction of atoms with the standing intensity wave (periodicity $\lambda/2 = 205$ nm) formed by counterpropagating spectroscopy laser beams. The simulation describes the atoms as delocalized over nodes and antinodes, crossing the standing wave with speed $v = 200$ m s$^{-1}$ and at angle $\delta\alpha + \alpha_0$ to orthogonal (transverse velocity of $v_x \approx v(\delta\alpha + \alpha_0)$; Fig. 2b). Results are shown for 2S–6P$_{1/2}$ (solid blue line) and 2S–6P$_{3/2}$ (dashed red line) transitions (at equal Rabi frequency), for which, respectively, 3.9% and 7.9% of the 6P level decays lead back to the 2S level (Fig. 2a). The hypothetical situation without back decay is also shown (dotted black line). At $v_x = v_{rec} \approx 0.97$ m s$^{-1}$ (the recoil velocity), a Bragg resonance occurs, leading to a large shift. The shift is symmetric in $v_x$ and negative (positive) for $|v_x| \lesssim v_{rec}$ ($|v_x| \gtrsim v_{rec}$). The simulated distribution (smoothed and scaled for visibility) of $v_x$ for an atomic beam offset angle $\alpha_0$ of 0 mrad (12 mrad) is indicated at the top (bottom) for velocity groups $\tau_2$, $\tau_{13}$ and $\tau_{16}$ (arbitrary vertical units). **b**, Detail of the Bragg resonance. **c**, Similar to Fig. 2c but for a line scan with $\alpha_0 = 12$ mrad, which splits the line into two Doppler components. The line is fitted with Voigt doublet line shape (Methods). **d**, Doppler-free transition frequency $\nu_e$ (blue circles) of the 2S–6P$_{1/2}$ transition for data taken at $P_{2S-6P} = 30$ μW and $|\alpha_0| = 0$, 7.5 and 12 mrad and corrected for all systematic effects except LFS. The simulation of LFS (orange circles) is in excellent agreement with the experimental data. Error bars show combined statistical and systematic uncertainty (experimental data) or uncertainty from input parameters (simulation).

Diffraction of matter waves, as for light waves, requires some degree of spatial coherence, with diffraction becoming important when the atoms' transverse coherence length $l_{c,t}$ along the standing wave is comparable with its periodicity of $\lambda/2 = 205$ nm. Treating the cryogenic nozzle as a thermal source of atoms, we find $l_{c,t}$ to be $\lambda_{dB,th} = 0.8$ nm (ref. 32) and therefore much smaller than $\lambda/2$, in which $\lambda_{dB,th} = \sqrt{h^2 k_B T_N / 2\pi m_H}$ is the thermal de Broglie wavelength ($m_H$, hydrogen mass; $k_B$, Boltzmann constant). However, as is well known from the van Cittert–Zernike theorem[33], the transverse coherence length is enhanced by propagation, as seen in matter wave interference of large molecules from a thermal source[34]. At a distance $L$ from the nozzle, this results in

$$l_{c,t} \approx (L/r_1)\lambda_{dB}/\pi \qquad (2)$$

for atoms with a speed $v$ and de Broglie wavelength of $\lambda_{dB} = h/m_H v$. For $v = 200$ m s$^{-1}$, the mean speed of atoms examined in the experiment, $l_{c,t}$ is 129 nm at the standing wave and hence comparable with its periodicity. We therefore need to model the atoms as partially coherent matter waves, that is, the atoms are partially delocalized over the standing wave, unlike the localized description[35] valid for different experimental conditions[28,29].

We use the Wigner function[36] to find a quantum-mechanical description of the atomic beam, quantizing the motion along the standing wave while treating other directions classically. This results in a comparable value of $l_{c,t}$ as estimated above[23]. The interaction of the atoms with the standing wave can be described in the combined basis of internal energy levels and external momenta along the standing wave (Supplementary Methods). The absorption of a photon from the forward-travelling (+) or backward-travelling (−) beam excites atoms from the 2S to the 6P level and changes their momentum along the beams by $\pm \hbar K_L$ (wavenumber $K_L = 2\pi/\lambda$; $\hbar = h/2\pi$). Likewise, stimulated emission into either beam returns atoms to the 2S level and changes their momentum by $\mp \hbar K_L$. This corresponds to a change in velocity by the recoil velocity, $v_{rec} = \hbar K_L / m_H \approx 0.97$ m s$^{-1}$, comparable with the typical transverse velocity $v_x$ of the atoms. Because the 6P level predominantly spontaneously decays to the 1S ground level, from which there is no re-excitation while the signal is recorded, the maximum relevant momentum change is here $\pm 4\hbar k$.

The small fraction $\gamma_{ei}/\Gamma$ of the spontaneous decays back to the initial 2S level leads to a random momentum change along the direction of the standing wave, as the direction of the emitted photon is randomly distributed[37]. The associated Doppler shift of such a decay (at most twice the recoil shift of $\Delta \nu_{rec} = 1{,}176.03$ kHz) is below the natural linewidth $\Gamma$ and therefore the atom can be re-excited to the 6P level and ultimately contribute to the signal. However, because $\gamma_{ei}/\Gamma \ll 1$, only at most one such back decay is relevant here.

We first simulate the interaction of the atoms with the standing wave for an atom initially in a momentum eigenstate, that is, as a fully delocalized matter wave, with transverse velocity $v_x \approx v(\delta\alpha + \alpha_0)$. Figure 3a shows the LFS $\nu_{\text{LFS}}$, found with a Voigt line shape fit to the simulated line shape, for an atom with $v = 200$ m s$^{-1}$ (note that $\nu_{\text{LFS}}$ is always symmetric in $v_x$). The hypothetical situation without back decay to the 2S level (dotted black line) describes pure diffraction of the matter wave on the standing wave. At $v_x = v_{\text{rec}}$, the Bragg condition is met and the atoms coherently scatter photons between the two counterpropagating beams, resulting in a narrow resonance with shifts exceeding 200 kHz (Fig. 3b). Below this resonance, the shift is negative and approximately constant around zero $v_x$, and above the resonance, the shift is positive and tends to zero as $v_x$ increases. Allowing back decay, on the other hand, allows scattering of the atoms on the standing wave. This primarily leads to a positive shift above the resonance that approximately scales with $\gamma_{\text{ei}}/\Gamma$, as shown for the 2S–6P$_{1/2}$ ($\gamma_{\text{ei}}/\Gamma = 3.9\%$; blue solid line) and 2S–6P$_{3/2}$ ($\gamma_{\text{ei}}/\Gamma = 7.9\%$; dashed red line) transitions.

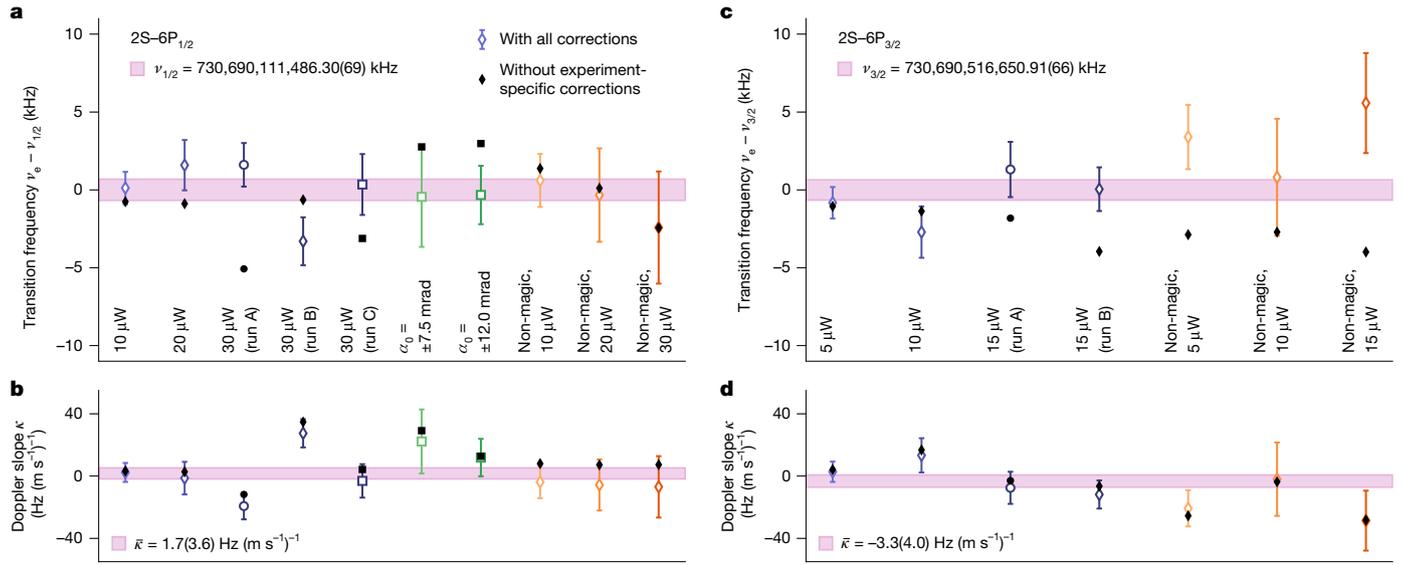

**Fig. 4 | Transition frequencies and Doppler slopes for the two examined 2S–6P transitions. a**, Measured transition frequencies $\nu_e$ of the 2S–6P$_{1/2}$ transition for different data groups (that is, combinations of experimental parameters and measurement runs), as labelled below each data point and listed in Extended Data Table 3. Data were taken for various combinations of spectroscopy laser power (10, 20 and 30 μW), atomic beam offset angle $\alpha_0$ ($\alpha_0 = 0$ mrad unless stated otherwise) and laser polarization angle $\theta_L$ (magic angle $\theta_L = 56.5°$ by default; non-magic angle $\theta_L = 146.5°$ marked accordingly). The data were collected over several months in 3-week-long runs, run A (circles), run B (diamonds) and run C (squares). Error bars show combined statistical and systematic one-standard-deviation ($\sigma$) uncertainty and the purple-shaded area shows the weighted mean of all data and its 1$\sigma$ uncertainty. Black markers show the results without experiment-specific corrections, including for the light force, QI and first-order and second-order Doppler shifts (Extended Data Table 4). **b**, Doppler slopes $\kappa$ for the data groups of panel **a**, determined in situ using velocity-resolved detection (Fig. 2). The mean speed of the atomic beam is $\langle \bar{v} \rangle \approx 195$ m s$^{-1}$. $\bar{\kappa}$ is the weighted mean Doppler slope. **c,d**, Same as panels **a** and **b** but for the 2S–6P$_{3/2}$ transition. Spectroscopy laser powers have a 2:1 ratio for the 2S–6P$_{1/2}$ and 2S–6P$_{3/2}$ transitions to keep their Rabi frequencies identical. The Pearson correlation coefficient between $\nu_e$ and $\kappa$ is $r = -0.78$ over all data groups.

The partially coherent atomic beam corresponds to an incoherent sum over atoms in various momentum eigenstates (Supplementary Methods). Because the transverse velocity spread of the beam is on the order of $v_{\text{rec}}$ (see shaded area at the top of Fig. 3a showing the velocity distribution of selected velocity groups and Extended Data Table 2), this prevents us from experimentally resolving the Bragg resonance and partly averages out the shift (to at most −0.96 kHz and −1.28 kHz for the 2S–6P$_{1/2}$ and 2S–6P$_{3/2}$ transitions, respectively, both occurring for velocity group $\tau_{13}$).

To test our model and simulation of the LFS, we measured the transition frequency of the 2S–6P$_{1/2}$ transition at atomic beam offset angles of $\alpha_0 = \pm 7.5$ mrad and $\alpha_0 = \pm 12$ mrad, as well as the dataset with $\alpha_0 = 0$ mrad (Fig. 3c,d). In particular, we find the difference in Doppler-free transition frequency between the data taken at $\alpha_0 = \pm 12$ mrad and $\alpha_0 = 0$ mrad (both for $P_{2S-6P} = 30$ μW), having applied a correction of 0.17(28) kHz for all systematic effects except the LFS, to be

$$\nu_e^{\text{(no LFS corr.)}}(\pm 12 \text{ mrad}) - \nu_e^{\text{(no LFS corr.)}}(0 \text{ mrad}) = 4.32(1.83) \text{ kHz}. \quad (3)$$

The uncertainty is dominated by the statistical uncertainty of 1.63 kHz of the ±12 mrad data. This difference is in excellent agreement with the LFS simulation, which predicts $\nu_{e,\text{LFS}}(\pm 12 \text{ mrad}) - \nu_{e,\text{LFS}}(0 \text{ mrad}) = 4.21(61)$ kHz. The accompanying distortion of the experimental line shape, clearly observable in the asymmetry of fit residuals (Extended Data Fig. 1), is distinctly different for the two values of $\alpha_0$ and in excellent agreement with the simulations. The measured LFS at $\alpha_0 = \pm 7.5$ mrad is also in agreement with the simulation but has a comparatively large statistical uncertainty of 3.04 kHz (Fig. 3d). This comparison is also a powerful test of the Doppler shift suppression scheme (equation (3) includes a Doppler shift correction of $\Delta\nu_e = -1.92(1.81)$ kHz) because of the large difference in $\alpha_0$ and of the data analysis because of the qualitatively different line shapes involved.

## Quantum interference

We use the QI simulations developed and extensively tested in ref. 2 to estimate any remaining QI shift, which we find to be −0.25(36) kHz and 0.05(15) kHz for the 2S–6P$_{1/2}$ and 2S–6P$_{3/2}$ transitions, respectively (Methods). These values include a small subset of data taken orthogonal to the magic angle, at which QI distortions are larger, to test our simulations. Also, we make use of the fact that the distortions are of opposite sign (and different magnitude) for the two transitions[22] and combine the 2S–6P$_{1/2}$ and 2S–6P$_{3/2}$ transition frequencies with a 1:2 ratio into the 2S–6P fine-structure centroid (Methods). This reduces the shift to only −0.05(2) kHz.

## 2S–6P transition frequencies

By averaging all available data (Fig. 4), we find the transition frequencies of the 2S–6P$_{1/2}$ transition ($\nu_{1/2}$) and the 2S–6P$_{3/2}$ transition ($\nu_{3/2}$) to be

# Article

$$v_{1/2} = 730{,}690{,}111{,}486.30(49)_{\text{stat}}(49)_{\text{sys}} \text{ kHz} \tag{4}$$
$$= 730{,}690{,}111{,}486.30(69) \text{ kHz } [0.94 \text{ ppt}],$$

$$v_{3/2} = 730{,}690{,}516{,}650.91(60)_{\text{stat}}(28)_{\text{sys}} \text{ kHz} \tag{5}$$
$$= 730{,}690{,}516{,}650.91(66) \text{ kHz } [0.90 \text{ ppt}].$$

The final one-standard-deviation uncertainties ($\sigma$) are the combined statistical ('stat') and systematic ('sys') uncertainties, with the former corresponding to the Doppler shift extrapolation uncertainty. All corrections and uncertainties are listed in Extended Data Table 4, with contributions not discussed in the main text detailed in Methods and Supplementary Methods. The results from the two detectors, averaged here, agree within their uncertainties. The average Doppler slopes $\bar{\kappa}$ of both transitions are compatible with zero.

By taking the difference of the two measured transition frequencies, we find the 6P fine-structure splitting between the $6P_{1/2}^{F=1}$ and $6P_{3/2}^{F=1}$ levels,

$$\Delta v_{\text{FS}}(6P) = v_{3/2} - v_{1/2} = 405{,}164.62(97) \text{ kHz}. \tag{6}$$

Extended Data Table 4 lists the corrections and uncertainties. The QED prediction $\Delta v_{\text{FS,QED}}(6P) = 405{,}164.51(1)$ kHz (Methods), which at our level of accuracy does not depend on $r_p$, is in excellent agreement ($\Delta v_{\text{FS}}(6P) - \Delta v_{\text{FS,QED}}(6P) = 0.11(97)$ kHz). This comparison tests the (uncorrelated) Doppler shift extrapolation and corrections for the light force, QI and dc-Stark shifts, with the last two of opposite sign for the two transitions.

Finally, we combine the two transition frequencies into the 2S–6P fine-structure centroid,

$$v_{2S-6P} = \frac{1}{3}v_{1/2} + \frac{2}{3}v_{3/2} + \Delta v_{\text{HFS}}(v_{2S-6P}) \tag{7}$$

$$v_{2S-6P} = 730{,}690{,}248{,}610.79(48) \text{ kHz } [0.66 \text{ ppt}], \tag{8}$$

with the hyperfine-structure (HFS) correction $\Delta v_{\text{HFS}}(v_{2S-6P}) = -132{,}985.25(1)$ kHz (Methods). The corrections and uncertainties for $v_{2S-6P}$ are listed in Table 1. The total applied correction, excluding the precisely known recoil shift and HFS corrections, corresponds to 3.6$\sigma$, with the largest individual correction (for the LFS) corresponding to 2.4$\sigma$. The relative uncertainty corresponds to a sixfold improvement over our previous 2S–4P measurement[2].

## $R_\infty$ and $r_p$ from atomic hydrogen

Combining the 2S–6P fine-structure centroid $v_{2S-6P}$ of equation (8) with a measurement of the 1S–2S transition frequency[18], and using equation (1), we determine the proton rms charge radius as

$$r_p = 0.8406(5)_{\text{QED}}(14)_{\text{exp}} \text{ fm} = 0.8406(15) \text{ fm}. \tag{9}$$

The total uncertainty arises mostly from the uncertainty of $v_{2S-6P}$ ('exp'), with the QED uncertainty of equation (1) ('QED') approximately threefold lower (contributions from the 1S–2S transition frequency and other physical constants are negligible). Figure 1 shows equation (9) with other relevant determinations of $r_p$. Equation (9) is the most precise value for $r_p$ from any measurement other than the muonic measurement[9,15], being 2.5-fold and sixfold more precise than the next best determination from atomic hydrogen[5] and our 2S–4P measurement[2], respectively. Equation (9) is in excellent agreement with the muonic measurement[15] ($r_p = 0.84060(39)$ fm) but disagrees with the CODATA 2014 proton radius[16] by 5.5$\sigma$. Instead of $r_p$, we may also determine the 1S Lamb shift as $\mathcal{L}_{\text{exp}}(1S) = 8{,}172{,}744.13(14)_{\text{QED}}(3.56)_{\text{exp}}$ kHz = $8{,}172{,}744.1(3.6)$ kHz (Methods).

**Table 1 | Corrections $\Delta v$ and uncertainties $\sigma$ for the determination of the 2S–6P fine-structure centroid $v_{2S-6P}$**

| Contribution | $\Delta v$ (kHz) | $\sigma$ (kHz) |
|---|---|---|
| First-order Doppler shift | 0.34 | 0.43 |
| – Extrapolation (statistical) | 0.34 | 0.43 |
| – Simulation of atom speeds | – | 0.01 |
| Simulation corrections | 1.05 | 0.17 |
| – LFS | 1.15 | 0.17 |
| – QI shift | 0.05 | 0.02 |
| – Second-order Doppler shift | −0.14 | 0.01 |
| dc-Stark shift | 0.05 | 0.07 |
| BBR-induced shift | 0.28 | 0.01 |
| Zeeman shift | 0.00 | 0.08 |
| Pressure shift | 0.00 | 0.02 |
| Sampling bias | 0.00 | 0.06 |
| Signal background | 0.00 | 0.03 |
| Laser spectrum | 0.00 | 0.07 |
| Frequency standard | 0.02 | 0.01 |
| Subtotal (experiment-specific contributions) | 1.74 | 0.48 |
| Recoil shift | −1,176.03 | 0.00 |
| HFS correction $\Delta v_{\text{HFS}}(v_{2S-6P})$ | −132,985.25 | 0.01 |
| Total (all contributions) | −134,159.54 | 0.48 |

All uncertainties correspond to one standard deviation. Indented entries detail subcontributions to the first-order Doppler shift and simulation corrections. The sum of the subcontributions may differ from the given total owing to rounding. BBR, blackbody radiation.

The Rydberg constant can be similarly extracted from the combination of the 1S–2S and 2S–6P transition frequencies, giving $R_\infty = 10{,}973{,}731.568152(14)$ m$^{-1}$ (1.3 ppt; Pearson correlation coefficient $r = 0.94$ with equation (9)). However, because the theory predictions of both frequencies depend on $R_\infty$ and $r_p$, we cannot make full use of the relative precision of 0.66 ppt of $v_{2S-6P}$.

## Test of the SM prediction

Having confirmed the muonic value of $r_p$, we now explicitly compare the SM prediction $v_{2S-6P,\text{SM}}$ for the 2S–6P fine-structure centroid with the experimentally determined value $v_{2S-6P}$. Combining the 1S–2S transition frequency[18] and the muonic value of $r_p$ (ref. 15) with equation (1) (Methods), we find

$$v_{2S-6P,\text{SM}} = 730{,}690{,}248{,}610.79(18)_{\text{QED}}(14)_{r_p} \text{ kHz} \tag{10}$$
$$= 730{,}690{,}248{,}610.79(23) \text{ kHz } [0.31 \text{ ppt}].$$

The prediction is in excellent agreement with our experimental result with a difference of $v_{2S-6P} - v_{2S-6P,\text{SM}} = 0.00(53)$ kHz, corresponding to a test of the SM at a relative precision of 0.7 ppt. This precision is comparable with the test of the SM prediction of the electron magnetic moment, which is at present limited to 0.7 ppt by discrepant measurements of $\alpha$ (ref. 12). Likewise, we compare the 1S Lamb shift prediction $\mathcal{L}_{\text{QED}}(1S) = 8{,}172{,}744.1(1.3)_{\text{QED}}(1.0)_{r_p}$ kHz = $8{,}172{,}744.1(1.7)$ kHz with $\mathcal{L}_{\text{exp}}(1S)$, which, to our knowledge, is the most precise test (0.5 ppm) of bound-state QED corrections so far.

Extended Data Table 1 lists the QED corrections and their uncertainties. Our experimental uncertainty is comparable with the muonic and hadronic vacuum polarization corrections and reaches the level of three-photon corrections ($\propto \alpha^5$), matching the highest-order terms in the prediction of the electron magnetic moment[12]. Although the experimental uncertainty is 2.7-fold larger than the QED uncertainty (dominated by two-photon and radiative-recoil corrections), the latter is partly estimated by the expected size of uncalculated terms[1,38] as opposed to experimental tests. In fact, a recent recalculation of the two-photon self-energy[39] would shift $\mathcal{L}_{QED}(1S)$ on the order of the experimental $\sigma$ of $\mathcal{L}_{exp}(1S)$.

Equivalently, we can compare the values of $R_\infty$ determined from $\nu_{2S-6P}$ or the 1S–2S transition frequency, each in combination with the muonic value of $r_p$ and equation (1). For $\nu_{2S-6P}$, this gives

$$R_\infty = 10{,}973{,}731.5681524(25)_{QED}(72)_{exp}(19)_{r_p}\ m^{-1} \qquad (11)$$
$$= 10{,}973{,}731.5681524(79)\ m^{-1}\ [0.75\ ppt],$$

with the uncertainty limited by the precision of $\nu_{2S-6P}$. Using the 1S–2S transition frequency yields a compatible $R_\infty = 10{,}973{,}731.5681523(65)\ m^{-1}$ ($r_{1S-2S} = 0.40$ correlation with equation (11)), limited by the uncertainty of $\mathcal{L}_{QED}(1S)$. Equation (11) is compatible with and 50% more precise than the $R_\infty$ world average[1], which has an expanded uncertainty accounting for scatter in the input data. Adding the 2S–6P measurement will probably reduce scatter in future adjustments as less precise measurements are removed. This situation is illustrated in Extended Data Fig. 2. Our measurement can also be used to improve new physics constraints on weakly interacting bosons with masses in the keV range[40,41].

The techniques demonstrated here may be used with any 2S–$n$P transitions in atomic hydrogen and deuterium[42], with a precision matching or exceeding that of SM predictions feasible. Together with complementary approaches[43–47], we expect this to substantially advance bound-state QED tests.

## Online content

Any methods, additional references, Nature Portfolio reporting summaries, source data, extended data, supplementary information, acknowledgements, peer review information; details of author contributions and competing interests; and statements of data and code availability are available at https://doi.org/10.1038/s41586-026-10124-3.

# Article

## Methods

### Data acquisition

Here we give further relevant details on the experimental scheme and apparatus (both described in detail in ref. 23). The run time of the cryogenic atomic beam is limited to freezing cycles of approximately 2 h by the accumulation of frozen (molecular) hydrogen inside the nozzle, which is removed by heating the nozzle to room temperature. The vacuum ($2 \times 10^{-7}$ mbar, dominated by molecular hydrogen) inside the 2S–6P spectroscopy region is maintained by differential pumping with a cryopump to minimize pressure shifts[50–52], with the temperature of the apparatus allowed to equilibrate on each measurement day before collecting data.

The power of the linearly polarized, 243-nm 1S–2S preparation laser[53,54] is resonantly enhanced in an in-vacuum, standing-wave cavity to $P_{1S-2S} \approx 1$ W per direction (297 μm $1/e^2$ intensity waist radius). The observed 1S–2S transition linewidth is approximately 3 kHz (FWHM; atomic detuning, as opposed to laser detuning), limited by single-photon ionization of the 2S level[55]. Therefore, the $2S_{1/2}^{F=1}$ levels are only populated by Doppler-sensitive two-photon excitation, leading to a population of approximately $7 \times 10^{-7}$ in each sublevel relative to the population in the $2S_{1/2}^{F=0}$ level. The detuning of the preparation laser is set several times per freezing cycle by observing the 1S–2S transition. An equal-slit-width optical chopper running at 160 Hz periodically blocks the preparation laser (which sets delay time $\tau = 0$ μs) to enable the velocity-resolved detection. The channel electron multipliers are switched off with a fast high-voltage switch while the preparation laser is unblocked to prevent saturation from scattered 243-nm light.

By using a linearly polarized 2S–6P spectroscopy laser[53,54] and the $2S_{1/2}^{F=0}$ level as the initial level, only transitions to the $6P_{1/2}^{F=1}$ hyperfine level (2S–6P$_{1/2}$ transition) and the $6P_{3/2}^{F=1}$ hyperfine level (2S–6P$_{3/2}$ transition) are dipole-allowed, whereas the excitation of the $6P_{1/2}^{F=0}$ and $6P_{3/2}^{F=2}$ levels is prevented by angular momentum conservation (Fig. 2a). The line strengths ($\propto \mu^2$, in which $\mu$ is the dipole moment) of the 2S–6P$_{1/2}$ and 2S–6P$_{3/2}$ transitions have a 1:2 ratio. We use spectroscopy laser powers $P_{2S-6P}$ with a ratio of 2:1 for the two transitions to keep their Rabi frequencies $\Omega_0 \propto \mu \sqrt{P_{2S-6P}}$ identical (the peak Rabi frequency is $(2\pi \times 126)$ krad s$^{-1}$ at our highest spectroscopy laser powers of 30 and 15 μW, respectively). Most 6P decays (branching ratio $\Gamma_{e-1S}/\Gamma = 88.2\%$) are Lyman decays to the 1S manifold, with the most energetic, direct Lyman-ε decay ($\Gamma_{det}/\Gamma = 80.5\%$) dominating, whereas the remaining $\gamma_{e-2S}/\Gamma = 11.8\%$ are Balmer decays to the 2S manifold, of which in turn a fraction $\gamma_{ei}/\Gamma$ leads back to the initial $2S_{1/2}^{F=0}$ level (see Section 1.2 in the Supplementary Methods). The metastable 2S levels are treated as stable here, as their natural lifetime (122 ms) is much longer than the time the atoms spend in the atomic beam (see Section 2.6 in the Supplementary Methods for the 2S decay contribution to the signal background).

The channel electron multipliers count the fluorescence photons from the 6P decays, either by detecting the photoelectrons emitted by the photons from the detector cylinder walls or, to a lesser extent, directly detecting the photons. Because the photoelectron yield strongly increases with photon energy (for both colloidal graphite and oxidized aluminium, the surface materials of the Faraday cage and the detector cylinder, respectively), fluorescence from Lyman-ε decay (13.2 eV photon energy) constitutes approximately 97% of the signal detected by the channel electron multipliers. The counts are binned into the 16 velocity groups by their delay time $\tau$, with the bins chosen to cover a wide range of mean speeds $\bar{v}$ while exhibiting a sufficient signal-to-noise ratio (bin width 50–550 μs; Extended Data Table 2). For each line scan, we accumulate counts over 160 chopper cycles at each spectroscopy laser detuning. Intermittently, excess scatter and spiking was observed for the bottom detector and its signal was subsequently discarded (≈11% of line scans; Extended Data Table 3). We attribute this to the bottom detector being cooled down to close to, and possibly below, its lower operating temperature limit because of its vicinity to the cryopump.

At least once per measurement day, the nozzle and the collimating aperture are centred on the preparation laser. At the start of each freezing cycle, the atomic beam offset angle $\alpha_0$, which is controlled by a linear motor equipped with a position sensor, is aligned to zero with a 1 mrad alignment uncertainty. This is achieved by blocking the returning beam of the spectroscopy laser (using an in-vacuum shutter in front of the high-reflectivity mirror of the AFR[25,26]), determining the (now unsuppressed) Doppler slope $\kappa$ for several angles and moving to the angle at which $\kappa$ is zero, which is set as $\alpha_0 = 0$ mrad. To record line scans at a non-zero angle $\pm\alpha_0$, we first move to either $+\alpha_0$ or $-\alpha_0$ (chosen randomly) and then to the opposite sign, recording typically 5–10 scans at each angle, and repeating this procedure several times per freezing cycle. The fibre–collimator distance of the AFR is optimized[25,26] at least once per freezing cycle.

We use fixed sets of 30 symmetric (15 unique) detunings Δ of the spectroscopy laser frequency to sample the 2S–6P fluorescence line shape, with different sets used for $\alpha_0 = 0, \pm 7.5$ and $\pm 12$ mrad to account for the different line shapes (see Section 5.1.3 and Table 5.2 of ref. 23), all of which have ±50 MHz as the largest detuning. The detunings were chosen to minimize the statistical uncertainty of the Doppler-free transition frequency $\nu_e$, whereas the number of detunings and the acquisition time (1 s) at each detuning were chosen to balance between sufficient line sampling, signal-to-noise ratio and number of line scans per freezing cycle. For each line scan, the order of the detunings is randomized to minimize the influence of drifts in the signal. At the beginning of each freezing cycle, the centre laser frequency (to which the detunings are added) was chosen randomly from a normal distribution. This distribution was centred on the 2S–6P transition frequency expected from the muonic measurement of the proton radius[9], with a standard deviation of 12 kHz to cover the transition frequency expected from the CODATA 2014 value of the proton radius[16]. The laser frequencies are referenced to the caesium frequency standard using an optical frequency comb[56,57] and a global navigation satellite system (GNSS)-referenced hydrogen maser (see Section 2.7 in the Supplementary Methods).

We switched between examining the 2S–6P$_{1/2}$ and 2S–6P$_{3/2}$ transitions several times during the measurement, except during measurement run C, for which only the 2S–6P$_{1/2}$ transition was examined for several values of $\alpha_0$ (Extended Data Table 3).

### Voigt and Voigt doublet line shapes

A Voigt line shape[23,58] is the convolution of a Lorentzian line shape (with FWHM linewidth $\Gamma_L$) and a Gaussian line shape (with FWHM linewidth $\Gamma_G$). It has a combined FWHM linewidth $\Gamma_F \approx 0.5346\Gamma_L + \sqrt{0.2166\Gamma_L^2 + \Gamma_G^2}$ (ref. 59) and amplitude $A$ and is centred on the resonance frequency $\nu_0$. A constant offset $y_0$ is added to account for the experimentally observed signal background, resulting in five free parameters.

The Voigt doublet line shape is here defined as the sum of two Voigt line shapes with generally different resonance frequencies $\nu_1$ and $\nu_2$ and amplitudes $A_1$ and $A_2$ but equal Lorentzian and Gaussian linewidths[23] (and a constant offset $y_0$), resulting in seven free parameters. Its resonance frequency is the centre of mass of the constituent line shapes, that is, $\nu_0 = (A_1\nu_1 + A_2\nu_2)/(A_1 + A_2)$.

### Data analysis

The resonance frequency $\nu_0$ of each velocity group and detector of each line scan is determined by least-square fitting Voigt line shapes to the signal (for data with $\alpha_0 = 0$ mrad, that is, data groups G1A–G12). For data with $\alpha_0 \neq 0$ mrad (data groups G13 and G14), in which the line can split into two Doppler components, the Voigt doublet line shape is used instead. For the fits, we assume the uncertainty on the signal $y_j$, that is, the number of fluorescence photons detected at detuning $j$, to be the photon-number shot (Poissonian) noise, given by $\sqrt{y_j}$

(because $y_j \gg 1$, the Poisson distribution can be approximated with a normal distribution).

We use reduced chi-squared $\chi^2_{red} = \chi^2/k$ (with $k$ degrees of freedom) as a measure of the goodness of fit. The average $\chi^2_{red}$ is expected to be 1 if the fitted line shapes exactly match the experimental line shapes and if the signal fluctuations are fully described by shot noise. We first discuss the Voigt fits ($k = 25$), for which we find $\chi^2_{red}$ to be substantially larger than 1 (up to $\chi^2_{red} \approx 4.6$ at our highest spectroscopy laser power) for all but the slowest velocity groups. Only for those velocity groups, which have a comparatively lower signal and Doppler broadening, does $\chi^2_{red}$ approach 1 ($\chi^2_{red} \approx 1.1$ for $\tau_{16}$ for all data groups).

We identify two distinct contributions to the increased $\chi^2_{red}$. First, there are small model deviations between the fitted and experimental line shapes, caused by non-Gaussian (and non-Lorentzian) broadening and saturation effects not included in the former and clearly observable in the fit residuals (up to approximately 2% deviation; see black circles at the top of Extended Data Fig. 1b). Our simulated line shapes (both LFS and QI simulations) include these effects and consequently show much better agreement with the experimental line shapes (see dashed and solid lines in Extended Data Fig. 1b). We use this to determine the effect of the deviations on $\chi^2_{red}$ by repeating the line shape fits using simulated data instead of experimental data. In this Monte Carlo simulation, the appropriate line shape simulation is scaled and offset to match the experimental data of each line scan and then shot noise is added to the simulated signal. The simulated $\chi^2_{red}$ values reproduce, and therefore the deviations explain, approximately 70% of the excess (that is, $\chi^2_{red} - 1$) in the experimental $\chi^2_{red}$ values.

Second, there are fluctuations in the signal that lead to excess (technical) noise above shot noise, which we attribute to fluctuations in the atomic flux (of metastable 2S atoms). In particular, we identify correlations between $\chi^2_{red}$ of a line scan and fluctuations of the nozzle temperature (Pearson correlation coefficient $r = 0.45$) and the preparation laser power ($r = 0.26$) during the line scan, both of which directly affect the number of 2S atoms reaching the spectroscopy region. The correlation with the spectroscopy laser power is much smaller ($r = 0.03$). We find that the simple assumption of a velocity-group-independent, 1% rms fluctuation of the signal from detuning to detuning explains, on average, the remaining excess of the experimental $\chi^2_{red}$ values.

The $\chi^2_{red}$ behaviour is very similar for the Voigt doublet fits ($k = 23$) to the $\alpha_0 = \pm 7.5$ mrad data (data group G13). For the $\alpha_0 = \pm 12$ mrad data (data group G14), the two Doppler components start to separate, particularly for fast velocity groups (Fig. 3c). This leads to large model deviations (up to 14%; see bottom of Extended Data Fig. 1b) from saturation effects, as some atoms interact with both spectroscopy laser beams ($|\Delta| \lesssim \Gamma_{2P}$), whereas others only interact with one beam ($|\Delta| \gtrsim \Gamma_{2P}$). Consequently, $\chi^2_{red}$ can exceed 20, which is largely explained (90% of excess) by the deviations, as again determined from the Monte Carlo simulation.

Asymmetric (about the line centre) deviations between the fitted and experimental line shapes, arising as a result of LFS and QI, do not substantially influence $\chi^2_{red}$ because of their small size. However, both symmetric and asymmetric deviations lead to a sampling bias, as discussed in Section 2.5 in the Supplementary Methods.

The uncertainty of the experimental resonance frequency $\nu_0$ is estimated from the line shape fits assuming only shot noise, that is, the technical noise is not taken into account at this point. The value of $\nu_0$ for each velocity group and detector of each line scan is corrected for the LFS and QI shift by subtracting, respectively, $\nu_{LFS}$ and $\nu_{QI}$, which are determined from the corresponding simulated line shapes (see Section 1.1 in the Supplementary Methods). The mean speed $\bar{\upsilon}$ and rms speed $\bar{\upsilon}_{rms}$ of each velocity group are also determined from these simulated line shapes (see Section 1.1 in the Supplementary Methods; Extended Data Table 2 gives the average value of $\bar{\upsilon}$ for each velocity group). The second-order Doppler shift $\Delta\nu_{SOD}$ is calculated using $\bar{\upsilon}_{rms}$ (see Section 2.1 in the Supplementary Methods) and likewise subtracted from $\nu_0$.

Next, using the simulated values of the speed $\bar{\upsilon}$, the Doppler shift extrapolation is performed to find the Doppler-free transition frequency $\nu_e$ and the Doppler slope $\kappa$. The values $\nu_e$ and $\kappa$ found for each line scan are inherently strongly correlated (Pearson correlation coefficient $r_e \approx -0.97$) and their uncertainties are found by propagating the uncertainties of $\nu_0$. Although the averaging process outlined below reduces this correlation, the correlation remains substantial ($r = -0.78$) between the averaged values of $\nu_e$ and $\kappa$ of different data groups. The LFS, QI shift and second-order Doppler shift all depend on the speed of the 2S atoms, either indirectly through the interaction time or directly. This results in non-zero Doppler slopes if not corrected for. In particular, the LFS, on average, would lead to $\kappa \approx -5$ Hz (m s$^{-1}$)$^{-1}$ and it is only by accounting for it that the experimentally determined values of $\kappa$ are, on average, compatible with zero.

The $\chi^2_{red}$ of the Doppler shift extrapolation ($k = 14$) is close to 1 ($\chi^2_{red} = 1.04(2)$ on average; standard deviation over data groups in parentheses), showing that the data are well described by a linear model and technical noise is small compared with shot noise. This contrasts with the excess noise observed in the line shape fits, as discussed above. We attribute this to the different timescales involved: the different velocity groups are recorded for about 100 µs and within 2,560 µs of each other, whereas the signal at each detuning is accumulated for 1 s before moving to the next detuning. That is, the Doppler shift extrapolation is mainly susceptible to technical noise on timescales of 100 µs, whereas the signal at different detunings is susceptible to technical noise on timescales of about 1 s, which is, for example, the timescale expected for nozzle temperature fluctuations.

Next we find the weighted mean of $\nu_e$ from the two detectors for each line scan (except when no data from the bottom detector are available, in which case data from the top detector are used), taking into account experimentally determined correlations ($r = 0.36(15)$ on average). We attribute these correlations, which tend to increase with spectroscopy laser power and therefore the signal, again to the detuning-to-detuning fluctuations in the atomic flux that are common mode to both detectors (and the velocity groups). We then form the weighted mean of the detector-averaged $\nu_e$ for each freezing cycle in a given data group. The reduced chi-squared $\chi^2_{red,FC}$ of this average is typically greater than 1 ($\overline{\chi}^2_{red,FC} = 1.44(23)$ on average), which we attribute again to fluctuations in the atomic flux. We account for this excess scatter by scaling the uncertainty of $\nu_e$ by the corresponding $\sqrt{\chi^2_{red,FC}}$ if $\chi^2_{red,FC} > 1$. This procedure shifts the determined transition frequencies (by changing the weighting of the data) by less than 20 Hz, much smaller than the associated uncertainties. A weighted mean of the detector-averaged, uncertainty-scaled $\nu_e$ is formed for each data group and the remaining corrections are applied (Table 1 and Extended Data Table 4). Finally, the relevant data groups are (weighted) averaged to find the transition frequencies of the 2S–6P$_{1/2}$ and 2S–6P$_{3/2}$ transitions. Throughout the analysis, the weights of the averages are based only on the (scaled) statistical frequency uncertainty (including for the Doppler slope $\kappa$), with the (correlated) uncertainty of the corrections not included in the weights. The statistical weights $w_{2S-6P}$ of the data groups for the determination of the 2S–6P fine-structure centroid $\nu_{2S-6P}$ are given in Extended Data Table 3.

The data analysis was blinded by adding a randomly chosen offset frequency to the transition frequencies. The offset frequency was only removed after all of the main systematic effects had been studied. Further small corrections, identified after the offset frequency was removed, resulted in a negligible shift of the determined transition frequencies of at most 10 Hz. The results of the data analysis (performed by L.M.) were confirmed by a second, independently implemented analysis (performed by V.W.).

### Modelling of atomic beam and fluorescence line shape

The fluorescence line shape of the 2S–6P transition is modelled by a Monte Carlo simulation of the atomic beam as a set of atomic trajectories, the trajectories' interaction with the 1S–2S preparation and 2S–6P spectroscopy lasers and their contribution to the fluorescence signal. The procedure is described in detail in Section 1.1 in the Supplementary Methods. Two complementary models describe the interaction with the spectroscopy laser, the QI model (see below and Section 1.2 in the Supplementary Methods) and the LFS model (see main text and Section 1.3 in the Supplementary Methods).

**Speed distribution of atomic beam**

We use the signal of the velocity groups as a time-of-flight measurement of the speed distribution of the atomic beam. To this end, we compare the experimental values of the line amplitudes $A$ of the velocity groups to the values of $A$ of line shapes simulated using a given speed distribution (see Section 1.1 in the Supplementary Methods). We find that the probability distribution of the speed $v$ of atoms leaving the nozzle towards the 2S–6P spectroscopy region is well described by a modified Maxwell–Boltzmann flux distribution for a wide range of experimental parameters (see ref. 23 for details). The flux distribution is given by

$$p_{\text{eff}}(v) = \mathcal{N} v^3 e^{-\frac{m_H v^2}{2 k_B T_N}} e^{-\frac{v_{\text{cut-off}}}{v}}, \quad (12)$$

in which $\mathcal{N}$ is a normalization constant. The extra factor $\exp(-v_{\text{cut-off}}/v)$ accounts for the depletion of slower atoms through collisions inside the nozzle, inside the beam and with the background gas[60–62]. A similar depletion has been observed in our 2S–4P measurement and other atomic hydrogen beams[3,63].

Using the above comparison of experimental data with simulations, $v_{\text{cut-off}}$ is found to be 50 m s$^{-1}$ on average. Extended Data Table 3 lists the average value and variation for each data group. A substantial part of the variation is because of the fact that $v_{\text{cut-off}}$ typically increases during a freezing cycle (see Fig. 6.1 of ref. 23). This is because the accumulation of frozen hydrogen decreases the diameter of the nozzle, causing an increase in the gas pressure and therefore collisions inside the nozzle. We take this variation into account when determining (the uncertainty of) the mean speeds of the velocity groups and the simulation corrections (see Section 1.1 in the Supplementary Methods and Extended Data Table 5).

**QI shift**

We simulate QI-distorted line shapes with a model combining optical Bloch equations with simulations of the spatial detection efficiency, averaged over a set of trajectories representing the atomic beam (see Sections 1.1 and 1.2 in the Supplementary Methods). The QI shift is here defined as the centre frequency of a line shape fit to the simulated line shapes (see Section 1.1 in the Supplementary Methods). The validity of this approach was demonstrated by the excellent agreement between the observed and simulated QI shifts in our previous measurement of the 2S–4P transition[2], in which the shifts were more than sevenfold larger because of the smaller detection solid angle and larger linewidth[22].

The simulated QI shifts, as a function of polarization angle $\theta_L$, are found to be at most 7.1 kHz and −4.1 kHz for the 2S–6P$_{1/2}$ and 2S–6P$_{3/2}$ transitions at our highest spectroscopy laser power (used to give upper limits here and below), respectively. At the magic angle $\theta_L = 56.5°$, the shifts reduce to at most −0.87(54) kHz and 0.45(27) kHz (including ±3° alignment uncertainty; again at the highest spectroscopy laser power). When we account for data taken at $\theta_L = 146.5°$ (see below) and at lower laser powers (with statistical weights as given in Extended Data Table 3), we obtain the overall simulated shifts (−0.25(36) kHz and 0.05(15) kHz) given in the main text. The cancellation inherent in the 2S–6P fine-structure centroid reduces the shift to at most −0.37 kHz at any polarization angle and to below 0.01 kHz at around $\theta_L = 56.5°$ (below the otherwise negligible ac-Stark shift; see Section 1.2 in the Supplementary Methods). Including all data results in the shift of −0.05(2) kHz given in the main text.

The magic angle $\theta_L = 56.5°$ used in the measurement was determined with simulations before the measurement began, whereas more refined simulations of the spatial detection efficiency (completed after the measurement and used for all simulation results given here) result in a magic angle of approximately 52°. Moreover, the magic angle also slightly shifts with laser power (by up to 2° for the powers used here; see Section 1.2 in the Supplementary Methods). However, the QI shifts are still strongly suppressed at the magic angle used, despite it being slightly different from the optimal value.

To test our QI model and simulations, a limited amount of data were taken at $\theta_L = 146.5°$ (Fig. 4 and Extended Data Table 3), that is, orthogonal to the magic polarization angle, which we compare with the data taken at $\theta_L = 56.5°$. For the 2S–6P$_{1/2}$ transition, the difference in Doppler-free transition frequency of $v_e(\theta_L = 146.5°) - v_e(\theta_L = 56.5°) = -0.01(1.69)$ kHz is in excellent agreement with 0 after correcting for a differential QI shift of 3.44(92) kHz and a differential Doppler shift of −1.42(1.42) kHz. We may also compare the experimental and simulated line shape distortions at $\theta_L = 146.5°$, in which the QI shift dominates over the LFS, by comparing the asymmetry of the experimental and simulated fit residuals (Extended Data Fig. 1). We find excellent agreement, especially at detunings larger than the linewidth, for which the line shape distortions from QI are largest.

For the 2S–6P$_{3/2}$ transition, we find a moderate (2.3 standard deviations) tension with a difference of $v_e(\theta_L = 146.5°) - v_e(\theta_L = 56.5°) = 4.08(1.77)$ kHz in the transition frequency, having corrected for a differential QI shift of −1.78(47) kHz. The removed differential Doppler shift is likewise significantly non-zero (−3.66(1.68) kHz). This correlation is consistent with, but not conclusive evidence for, the non-zero difference being caused by random errors affecting the Doppler shift extrapolation. This conclusion is also supported by the fact that there is no tension in the velocity-group-averaged resonance frequency, that is, assuming zero Doppler shift (see Fig. 6.9 of ref. 23). Furthermore, we find the line shape distortions at $\theta_L = 146.5°$ to be compatible with our QI simulations but incompatible with a QI shift of opposite sign and twice the magnitude as implied by the measured difference (Extended Data Fig. 1). We therefore conclude that the tension probably results from random errors in the determination of the resonance frequencies.

**2S–6P fine-structure centroid, 6P fine-structure splitting and HFS corrections**

The 2S–6P fine-structure centroid $v_{2S-6P}$ is the transition frequency from the 2S HFS centroid to the 6P fine-structure centroid. It is determined from the two measured transition frequencies $v_{1/2}$ and $v_{3/2}$ for the transitions from the $2S_{1/2}^{F=0}$ level to the $6P_{1/2}^{F=1}$ level (2S–6P$_{1/2}$ transition) and the $6P_{3/2}^{F=1}$ level (2S–6P$_{3/2}$ transition), respectively, by first correcting $v_{1/2}$ and $v_{3/2}$ for the 2S and 6P HFS and then averaging the corrected $v_{1/2}$ and $v_{3/2}$ weighted by their fine-structure multiplicity ratio of 1:2 (equivalent to the ratio of the line strengths $\propto \mu^2$ of the 2S–6P$_{1/2}$ and 2S–6P$_{3/2}$ transitions, in which $\mu$ is the dipole moment) to find the 6P fine-structure centroid. This results in equation (7), with the HFS corrections included in $\Delta v_{\text{HFS}}(v_{2S-6P})$, as detailed below.

The hyperfine interaction splits the fine-structure levels 2S$_{1/2}$, 6P$_{1/2}$ and 6P$_{3/2}$ into doublets[64,65]. The relevant HFS levels $2S_{1/2}^{F=0}$, $6P_{1/2}^{F=1}$ and $6P_{3/2}^{F=1}$ are shifted from the fine-structure levels by the HFS energies $\Delta v_{\text{HFS}}(2S_{1/2}^{F=0})$, $\Delta v_{\text{HFS}}(6P_{1/2}^{F=1})$ and $\Delta v_{\text{HFS}}(6P_{3/2}^{F=1})$, respectively (see Fig. 6.11 of ref. 23 for the relevant level scheme). The value of $\Delta v_{\text{HFS}}(2S_{1/2}^{F=0})$ and its uncertainty are obtained from a measurement[66] of the 2S HFS splitting $\Delta v_{\text{HFS}}(2S_{1/2})$ as

$$\Delta v_{\text{HFS}}(2S_{1/2}^{F=0}) = -(3/4)\Delta v_{\text{HFS}}(2S_{1/2}) = -133,167,625.7(5.0) \text{ Hz}. \quad (13)$$

$\Delta v_{\text{HFS}}(6P_{1/2}^{F=1})$ and $\Delta v_{\text{HFS}}(6P_{3/2}^{F=1})$ can be calculated as detailed in refs. 64,65. They include small corrections from off-diagonal elements in the HFS Hamiltonian, which mix HFS levels with the same value of $F$ but different values of $J$. Because only the $F = 1$ level of each HFS doublet is shifted by this effect, the centres of gravity of the HFS doublets are shifted by $\Delta\nu_{\rm HFS}^{\rm o.d.}(6P_{1/2})$ and $\Delta\nu_{\rm HFS}^{\rm o.d.}(6P_{3/2})$. Using the values for the 6P HFS splittings $\Delta\nu_{\rm HFS}(6P_{1/2})$, $\Delta\nu_{\rm HFS}(6P_{3/2})$ and the values for $\Delta\nu_{\rm HFS}^{\rm o.d.}(6P_{1/2})$, $\Delta\nu_{\rm HFS}^{\rm o.d.}(6P_{3/2})$ given in Table 1 of ref. 64 (see also equations (29) and (30) in ref. 65 and comments therein), we find

$$\Delta\nu_{\rm HFS}(6P_{1/2}^{F=1}) = (1/4)\Delta\nu_{\rm HFS}(6P_{1/2}) + \Delta\nu_{\rm HFS}^{\rm o.d.}(6P_{1/2}) = 547{,}798(6)\ {\rm Hz}, \quad (14)$$

$$\Delta\nu_{\rm HFS}(6P_{3/2}^{F=1}) = -(5/8)\Delta\nu_{\rm HFS}(6P_{3/2}) + \Delta\nu_{\rm HFS}^{\rm o.d.}(6P_{3/2})$$
$$= -547{,}460(6)\ {\rm Hz}. \quad (15)$$

$\Delta\nu_{\rm HFS}(6P_{1/2}^{F=1})$ and $\Delta\nu_{\rm HFS}(6P_{3/2}^{F=1})$ are assumed to be fully correlated.

The resulting HFS correction of the 2S–6P fine-structure centroid $\nu_{\rm 2S-6P}$ is

$$\Delta\nu_{\rm HFS}(\nu_{\rm 2S-6P}) = -(1/3)\Delta\nu_{\rm HFS}(6P_{1/2}^{F=1})$$
$$-(2/3)\Delta\nu_{\rm HFS}(6P_{3/2}^{F=1}) + \Delta\nu_{\rm HFS}(2S_{1/2}^{F=0}) \quad (16)$$
$$= -132{,}985{,}250(10)\ {\rm Hz},$$

in which we rounded to the nearest 10 Hz, as done for all corrections.

### SM predictions of transition frequencies and bound-state QED test

We find the SM prediction $\nu_{\rm 2S-6P,SM}$ of the 2S–6P fine-structure centroid using

$$\nu_{\rm 2S-6P,SM} = \frac{(1/3)E_{6,1,1/2} + (2/3)E_{6,1,3/2} - E_{2,0,1/2}}{E_{2,0,1/2} - E_{1,0,1/2}}\nu_{\rm 1S-2S}$$
$$= \frac{E(2S-6P)}{E(1S-2S)}\nu_{\rm 1S-2S}, \quad (17)$$

in which $E_{nlj}$ are the fine-structure level energies from equation (1) and $\nu_{\rm 1S-2S}$ is the measured frequency of the 1S–2S hyperfine centroid[18]. This parametrization removes the explicit dependence of $E_{nlj}$ on $R_\infty$. Using the muonic value of $r_{\rm p}$ (ref. 15), we find the value of $\nu_{\rm 2S-6P,SM}$ given in equation (10).

Extended Data Table 1 lists the contributions to $\nu_{\rm 2S-6P,SM}$. Along with the Dirac eigenvalue ($cR_\infty f_{nlj}^{\rm Dirac}$ of equation (1); equation (30) in ref. 1), we list the individual QED corrections ($cR_\infty f_{nlj}^{\rm QED}$ and $cR_\infty \delta_{l0}(C_{\rm NS}/n^3)\,r_{\rm p}^2$ of equation (1); equations (32)–(64) in ref. 1), with the sum of the corrections corresponding to the Lamb shift $\mathcal{L}(\mathcal{L}(nlj)$ for a single fine-structure level; $\mathcal{L}(2S-6P) = (1/3)\mathcal{L}(6,1,1/2) + (2/3)\mathcal{L}(6,1,3/2) - \mathcal{L}(2,0,1/2)$ for the 2S–6P fine-structure centroid). Extended Data Table 1 also gives the QED-only uncertainty, which we define as the uncertainty excluding contributions from $r_{\rm p}$, $\alpha$, $m_{\rm p}/m_{\rm e}$ and $\nu_{\rm 1S-2S}$. All listed QED corrections scale as $cR_\infty C/n^3$ in leading order for S-states, in which $cR_\infty C$ is the corresponding leading-order QED correction to the 1S level. This includes the (leading order, $\propto \alpha^2 \times \alpha^2$, with the first $\alpha^2$ factor absorbed in $R_\infty = \alpha^2 m_{\rm e} c/2h$ in equation (1)) nuclear size correction highlighted in equation (1), for which $C = C_{\rm NS} r_{\rm p}^2$. Therefore, the corrections between different S-states are correlated, which is taken into account in the uncertainty of $\nu_{\rm 2S-6P,SM}$. Non-S-states have generally much smaller corrections (for example, the fractional corrections are $1.3 \times 10^{-6}$ of the 2S binding energy but $4.4 \times 10^{-9}$ of the 6P binding energy) and, in particular, their nuclear size corrections are at present negligible (at most $7 \times 10^{-16}$ of the 6P binding energy). There is no correlation between the different QED corrections and we find the total QED-only uncertainty of 179 Hz by adding the uncertainties of the corrections in quadrature. Overall, $E(2S-6P)$ and $E(1S-2S)$ are highly correlated ($r = 0.995$ if only considering QED-only uncertainty).

The other dominant source of uncertainty for $\nu_{\rm 2S-6P,SM}$ is the muonic value of $r_{\rm p}$, which contributes 138 Hz to the uncertainty of $\nu_{\rm 2S-6P,SM}$. Although the uncertainty of $r_{\rm p}$ itself partly originates from QED corrections[15], the QED predictions for the energy levels of hydrogen and muonic hydrogen are uncorrelated at the present level of accuracy[1] and we therefore treat the QED-only uncertainty and the uncertainty from $r_{\rm p}$ as uncorrelated. The uncertainty contributions from $\alpha$ (1.7 Hz), $m_{\rm p}/m_{\rm e}$ (0.1 mHz) and $\nu_{\rm 1S-2S}$ (3 Hz) are negligible. In total, this results in an uncertainty of 226 Hz on $\nu_{\rm 2S-6P,SM}$.

It is instructive to find the sensitivity of a prediction $\nu_{\rm SM}$, derived in the same way as $\nu_{\rm 2S-6P,SM}$, to changes in $C$. Making use of the $1/n^3$ scaling of the QED corrections (and using the approximation $E_{nlj} \approx chR_\infty(-1/n^2 + \delta_{l0}C/n^3)$, that is, ignoring all non-leading-order corrections to $f_{nlj}^{\rm Dirac}$ and all QED corrections except $C$), we find

$$\frac{\partial}{\partial C}\left(\frac{\nu_{\rm SM}(n,n',\tilde{n},\tilde{n}')}{cR_\infty}\right) \approx \left(\frac{\delta_{l0}}{n^3} - \frac{\delta_{l'0}}{n'^3}\right) - \left(\frac{\delta_{\tilde{l}0}}{\tilde{n}^3} - \frac{\delta_{\tilde{l}'0}}{\tilde{n}'^3}\right)\frac{1/n^2 - 1/n'^2}{1/\tilde{n}^2 - 1/\tilde{n}'^2}, \quad (18)$$

in which $n', l' \to n, l$ is the transition to be predicted (for example, 2S–6P transition for $\nu_{\rm 2S-6P,SM}$) and $\tilde{n}', \tilde{l}' \to \tilde{n}, \tilde{l}$ is the measured transition used as input (for example, 1S–2S transition for $\nu_{\rm 2S-6P,SM}$). The sensitivity of $\nu_{\rm 2S-6P,SM}$ is $\partial(\nu_{\rm 2S-6P,SM}/cR_\infty)/\partial C = 0.134$. By contrast, using the 1S–3S transition instead of the 2S–6P transition leads to a 1.8-fold lower sensitivity ($\partial(\nu_{\rm SM}/cR_\infty)/\partial C = 0.074$) because the relative contribution of $C$ to the 1S–3S transition frequency is approximately twice as large as for the 2S–6P transition frequency. Combined with its 1.5-fold smaller uncertainty, $\nu_{\rm 2S-6P}$ therefore tests $C$ with 2.7-fold higher precision than the 1S–3S measurement[5]. Because the nuclear size correction $cR_\infty C_{\rm NS}\,r_{\rm p}^2/n^3$ scales as $1/n^3$ like the other bound-state QED corrections $cR_\infty C/n^3$, comparing $r_{\rm p}$ found from atomic hydrogen and from muonic hydrogen is a test of bound-state QED, as any missing or miscalculated terms in atomic hydrogen of the form $cR_\infty C/n^3$ would lead to a discrepancy. Furthermore, the precision with which $r_{\rm p}$ can be extracted from a given combination of measurements is therefore a direct measure of the precision of the implied QED test. Note that, because Extended Data Table 1 lists the corrections of $\nu_{\rm 2S-6P,SM}$, which are approximately $-cR_\infty C/2^3$, the sensitivity relative to the listed corrections is $-8 \times 0.134 = -1.072$. For example, artificially removing the hadronic vacuum polarization shifts $\nu_{\rm 2S-6P,SM}$ by $-1.072 \times 425.1$ Hz $= -0.46$ kHz, or approximately one experimental $\sigma$ of $\nu_{\rm 2S-6P}$.

We may also extract the 1S Lamb shift $\mathcal{L}(1S)$ by writing equation (1) as

$$E_{nlj} = chR_\infty f_{nlj}^{\rm Dirac} + \delta_{0l}\frac{h\mathcal{L}(1S)}{n^3} + h\delta\mathcal{L}(nlj), \quad (19)$$

in which $\mathcal{L}(1S) \equiv \mathcal{L}(1,0,1/2) = cR_\infty(f_{1,0,1/2}^{\rm QED} + C_{\rm NS}r_{\rm p}^2)$ and $\delta\mathcal{L}(nlj) = \mathcal{L}(nlj) - \delta_{0l}\mathcal{L}(1S)/n^3$ is the state-specific Lamb shift. By combining equation (19) for the 2S–6P and 1S–2S transitions and using $\nu_{\rm 2S-6P}$ and $\nu_{\rm 1S-2S}$ as inputs, we find $\mathcal{L}_{\rm exp}(1S) = 8{,}172{,}744.13(14)_{\rm QED}(3.56)_{\rm exp}$ kHz $= 8{,}172{,}744.1(3.6)$ kHz. Terms proportional to state-specific Lamb shifts contribute 224 MHz to $\mathcal{L}_{\rm exp}(1S)$. The uncertainty is dominated by the uncertainty of $\nu_{\rm 2S-6P}$ ('exp'), with the QED uncertainty of the state-specific Lamb shifts ('QED') being much smaller and contributions from $\nu_{\rm 1S-2S}$ (22 Hz) and physical constants (10 Hz from $\alpha$; $m_{\rm p}/m_{\rm e}$, $r_{\rm p}$, $R_\infty$ less than 0.1 Hz) negligible. As expected from the discussion above, $\mathcal{L}_{\rm exp}(1S)$ is 2.7-fold more precise than the next best determination using the 1S–3S measurement[5].

The corresponding QED prediction of the 1S Lamb shift is found by combining equation (1) with the muonic value of $r_{\rm p}$, giving $\mathcal{L}_{\rm QED}(1S) = 8{,}172{,}744.1(1.3)_{\rm QED}(1.0)_{r_{\rm p}}$ kHz $= 8{,}172{,}744.1(1.7)$ kHz. Its uncertainty is dominated by both QED uncertainty and the uncertainty of $r_{\rm p}$, whereas other sources are negligible (3 Hz from $\alpha$; $m_{\rm p}/m_{\rm e}$, $R_\infty$ less than 0.01 Hz). $\mathcal{L}_{\rm exp}(1S)$ and $\mathcal{L}_{\rm QED}(1S)$ are uncorrelated, as their respective QED uncertainties are uncorrelated[1]. They are in excellent agreement and test the 1S Lamb shift and thereby bound-state QED corrections to 0.5 ppm. Complementary tests of bound-state QED in strong electromagnetic fields with highly charged ions at present achieve a relative precision of $1 \times 10^{-4}$ but can be more sensitive to terms of high order in

# Article

($Z\alpha$) ($Z$ is the nuclear charge number; omitted elsewhere here because $Z = 1$ for atomic hydrogen)[47,67,68].

Finally, the QED prediction $\Delta\nu_{\text{FS,QED}}(6P)$ for the 6P fine-structure splitting between the $6P_{1/2}^{F=1}$ and $6P_{3/2}^{F=1}$ levels is (using equations (14) and (15))

$$\Delta\nu_{\text{FS,QED}}(6P) = E_{6,1,3/2}/h - E_{6,1,1/2}/h + \Delta\nu_{\text{HFS}}(6P_{3/2}^{F=1})$$
$$- \Delta\nu_{\text{HFS}}(6P_{1/2}^{F=1}) \quad (20)$$
$$= 405{,}164.51(1) \text{ kHz}.$$

The nuclear size corrections in $\Delta\nu_{\text{FS,QED}}(6P)$ are of order $\alpha^2 \times \alpha^4$ and $\alpha^2 \times \alpha^5$ and amount to 70 mHz, as the leading-order correction term $cR_\infty C_{\text{NS}} r_p^2 \propto \alpha^2 \times \alpha^2$ of equation (1) and corrections of order $\alpha^2 \times \alpha^3$ only apply to S-states[1].

## dc-Stark shift

Static stray electric fields in the 2S–6P spectroscopy region can lead to a dc-Stark shift of the observed transition frequency. Here the dc-Stark shift $\Delta\nu_{\text{dc}}$ is well described as quadratic in strength $E = |\mathbf{E}|$ of the electric field $\mathbf{E}$ in all relevant experimental regimes, that is,

$$\Delta\nu_{\text{dc}} = \beta_{\text{dc}} E^2, \quad (21)$$

in which $\beta_{\text{dc}}$ is the applicable quadratic dc-Stark shift coefficient. We distinguish two experimentally relevant field strength regimes: the stray-field regime ($E < 1$ V m$^{-1}$), which covers the range of stray electric fields present in the experiment, and the bias-field regime ($E = 10$–$45$ V m$^{-1}$), which covers the range of applied bias fields used to determine the stray electric fields. Notably, although equation (21) is found to approximately hold in either regime, the coefficient $\beta_{\text{dc}}$ may differ, as is the case for the 2S–$6P_{3/2}$ transition (Section 1.4 in the Supplementary Methods). The quadratic behaviour arises because the energy levels contributing to the net shift are well separated in energy from the perturbed level in either of our regimes. However, the shift of the involved levels between the regimes can lead to substantially different energy separations and thereby different values of $\beta_{\text{dc}}$.

The Faraday cage surrounding the spectroscopy region (Fig. 2b) shields it from external electric fields, including those used to draw the photoelectrons to the channel electron multipliers (whose input surfaces are held at 270 V; see Section 4.6.2 of ref. 23). The colloidal graphite coating on all surfaces of the Faraday cage suppresses stray electric fields from the surfaces themselves (from charged oxide layers, contact potentials from dissimilar conductors or local changes in the work function) by, ideally, forming a uniform conductive layer with a uniform work function (see Section 4.6.1 of ref. 23). However, effects such as imperfect shielding of external fields, imperfect graphite coating or temperature gradients leading to thermoelectric voltages or gradients in the work function can prevent the complete suppression of stray fields.

To address this, we measure the stray electric field by applying voltages to the six electrodes forming the Faraday cage (meshes at the top and bottom and four equal-sized segments of the cylinder wall) and using the atoms themselves as field sensors (see Section 4.6.7 of ref. 23), similar to approaches in refs. 5,69. Equal and opposite bias voltages are applied to opposing electrodes to create a bias electric field $\mathbf{E} = E_i \hat{\mathbf{i}}$ with strength $|E_i| = 10$–$45$ V m$^{-1}$ along the given direction ($i = x, y, z$, as defined in Fig. 2b). By measuring the shifted 2S–6P transition frequency $\nu_e(E_i)$ from a fit to the fluorescence line shape of several line scans with opposite-polarity values of $E_i$ and using the quadratic dependence $\nu_e(E_i) = \beta_{\text{dc},i}(E_i - \Delta E_i)^2 + \nu_e(E_i = 0 \text{ V m}^{-1})$, we determine the stray electric field component $\Delta E_i$ along the given direction. This measurement also yields experimental values of $\beta_{\text{dc},i}$ at the bias field strengths for each transition. Extended Data Fig. 3 shows examples of such stray field measurements for both the 2S–$6P_{1/2}$ and 2S–$6P_{3/2}$ transitions, along with simulation results (see Section 1.4 in the Supplementary Methods). On average, each measurement includes eight line scans with non-zero bias field and there are 98 and 21 measurements for the 2S–$6P_{1/2}$ and 2S–$6P_{3/2}$ transitions, respectively. The stray electric field components determined during the three measurement runs are shown in Extended Data Fig. 4, in which we only include measurements using the 2S–$6P_{1/2}$ transition because using the 2S–$6P_{3/2}$ transition gives compatible, but substantially larger uncertainty, values of $\Delta E_i$. Overall, the stray electric field has a strength less than 1 V m$^{-1}$ and predominantly points along the axis of the detector cylinder (the $y$-direction).

Using these stray field measurements, we estimate the dc-Stark shift correction and its uncertainty for the transition frequencies. The determination of the relevant quadratic dc-Stark shift coefficients in the stray-field regime by the use of experimentally verified simulations is described in Section 1.4 in the Supplementary Methods. The weighted mean and standard deviation of each stray electric field component $\Delta E_i$ are determined over each of the three measurement runs, using the 2S–$6P_{1/2}$ stray field measurements, as shown in Extended Data Fig. 4. We treat each measurement run separately to account for differences in the detector assembly (the meshes in the detector cylinder were replaced between run A and run B; see Section 4.6.1 of ref. 23) and the month-long breaks in between the runs. Furthermore, we opt to use the standard deviation of the stray field components to estimate the dc-Stark shift uncertainty because we believe at least part of the variation of the stray fields to be physical in origin (as opposed to unaccounted excess measurement scatter) but are not confident that the resulting variations in the dc-Stark shift will average out. With this, we find dc-Stark shifts $\Delta\nu_{\text{dc}}$ of 0.20(21) kHz and –0.02(6) kHz for the 2S–$6P_{1/2}$ and 2S–$6P_{3/2}$ transitions, respectively, in which the uncertainty is dominated by the stray electric fields in the first case and by coefficients in the second case, and shifts for both transitions mainly determined by the dominant stray electric field component along the $y$-direction. $\nu_{1/2}$ and $\nu_{3/2}$ have been corrected for the dc-Stark shift by subtracting the corresponding value. The Pearson correlation coefficient of the correction between the 2S–$6P_{1/2}$ and 2S–$6P_{3/2}$ transitions is found to be $r = -0.30$ by propagation of uncertainty.

## Data availability

The experimental data and simulation results that support the findings of this study are available from Zenodo[70].

## Code availability

All code used for the data analysis and simulations of this study is available from the corresponding author.

**Acknowledgements** We thank S. G. Karshenboim for helpful discussions and W. Simon, K. Linner and H. Brückner for technical support. Computations were performed on the HPC systems at the Max Planck Computing and Data Facility. This work was supported by European Research Council (ERC) grant H-SPECTR (grant agreement ID 101141942). L.M. acknowledges a Feodor Lynen Fellowship from the Alexander von Humboldt Foundation. V.W. acknowledges support from the International Max Planck Research School for Advanced Photon Science (IMPRS-APS). A.M. and A.G. acknowledge support from the German Research Foundation (DFG) (project IDs 390524307 and EXC-2111-390814868, respectively). R.P. acknowledges support from the PRISMA+ Cluster of Excellence (EXC 2118/1) funded by the DFG in the German Excellence Strategy (project ID 39083149). T.W.H. acknowledges support from the Carl Friedrich von Siemens Foundation and the Max Planck Foundation.

**Author contributions** L.M. and V.W. set up, maintained and performed the experiment. L.M. wrote the software for and set up the data acquisition system. L.M. conducted the data analysis, which V.W. independently reproduced. A.M. and L.M. performed modelling and simulations. T.U., L.M. and V.W. evaluated the QED predictions and tests. L.M. prepared the manuscript. All authors (L.M., V.W., A.M., A.G., R.P., T.W.H. and T.U.) contributed to the discussion and analysis of the systematic uncertainties and edited the manuscript.

**Funding** Open access funding provided by Max Planck Society.

**Competing interests** The authors declare no competing interests.

**Additional information**
**Supplementary information** The online version contains supplementary material available at https://doi.org/10.1038/s41586-026-10124-3.
**Correspondence and requests for materials** should be addressed to Lothar Maisenbacher.
**Peer review information** *Nature* thanks Saïda Guellati-Khelifa and the other, anonymous, reviewer(s) for their contribution to the peer review of this work. Peer reviewer reports are available.
**Reprints and permissions information** is available at http://www.nature.com/reprints.

# Article

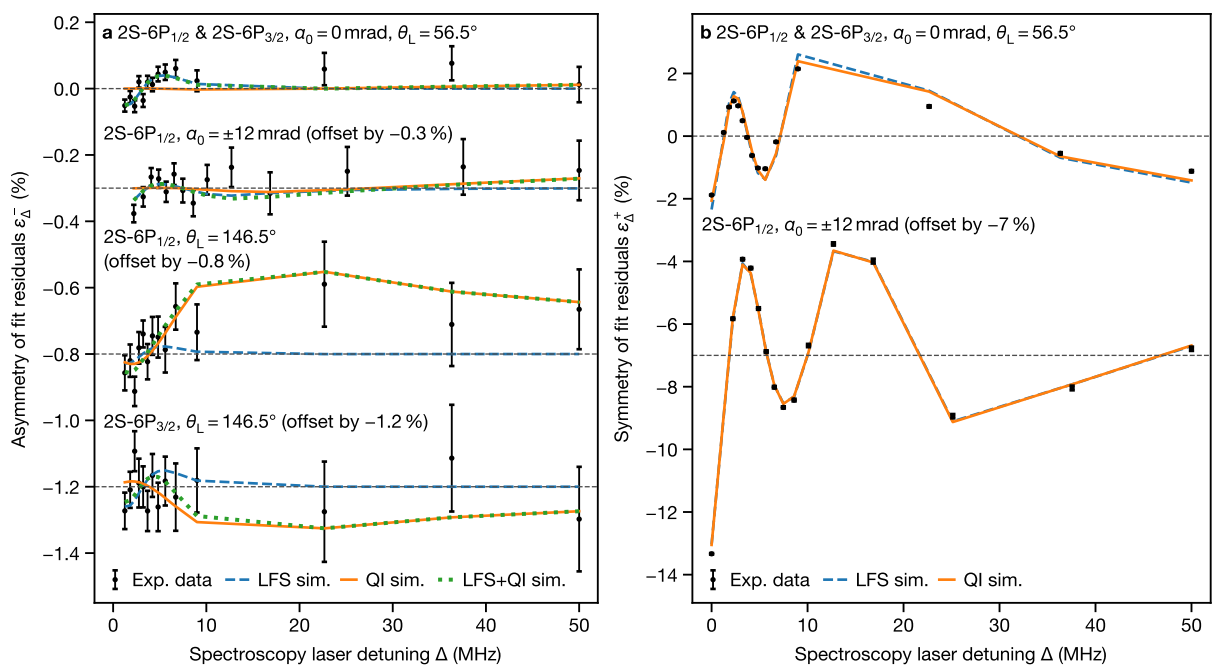

**Extended Data Fig. 1 | Test of line shape models. a**, Detuning asymmetry of fit residuals $\varepsilon_\Delta^-$ at spectroscopy laser detuning $\Delta$ for four different experimental settings, with experimental data (black circles), LFS simulation (dashed blue line) and QI simulation (solid orange line) shown. The sum of $\varepsilon_\Delta^-$ for LFS and QI simulations is also shown (dotted green line). The normalized fit residuals $\varepsilon_\Delta$ are given by $(y_\Delta - y_{\text{fit},\Delta})/y_{\text{fit},\Delta}$, in which $y_\Delta$ is the experimental or simulated signal and $y_{\text{fit},\Delta}$ the corresponding value from the Voigt or Voigt doublet line shape fit to the signal, and their detuning asymmetry and symmetry are defined as $\varepsilon_\Delta^- = (\varepsilon_\Delta - \varepsilon_{-\Delta})/2$ and $\varepsilon_\Delta^+ = (\varepsilon_\Delta + \varepsilon_{-\Delta})/2$. Note that the curves have been offset (dashed grey horizontal lines) to enhance visibility. Top, for data taken at zero atomic beam offset angle $\alpha_0$ and at magic polarization angle $\theta_L = 56.5°$, in which the QI effect is strongly suppressed, the experimentally observed asymmetry $\varepsilon_\Delta^-$ is well explained by the LFS simulation. All relevant data from measurement run B are used. See Fig. 2c for an example line shape. Second from top, for $\alpha_0 = \pm 12$ mrad (and magic polarization angle; see Fig. 3c for an example line shape), the LFS simulation predicts a different asymmetry compared with zero $\alpha_0$, which again matches the experiment well. This difference in asymmetry is directly observed in the transition frequency as shown in Fig. 3d. Third from top, for a non-magic polarization angle of $\theta_L = 146.5°$, the QI effect leads to a strong asymmetry at large detunings $\Delta \gtrsim 5$ MHz $\approx \Gamma$, which is observed in the experimental data (shown for the 2S–6P$_{1/2}$ transition; $\alpha_0 = 0$ mrad). At smaller detunings, both the QI effect and LFS contribute to the asymmetry and the sum of their asymmetries describes the experimental data well. Bottom, for the non-magic polarization angle for the 2S–6P$_{3/2}$ transition, the asymmetry from the QI effect has the opposite sign and is about twofold smaller compared with the 2S–6P$_{1/2}$ transition. **b**, Detuning symmetry of fit residuals $\varepsilon_\Delta^+$ for the first two cases of panel **a**. The deviations are caused by non-Gaussian broadening and saturation effects, which are not included in the fitted line shape models but are included in the simulations. Only results for the top detector are shown, as the detector signals are highly correlated.

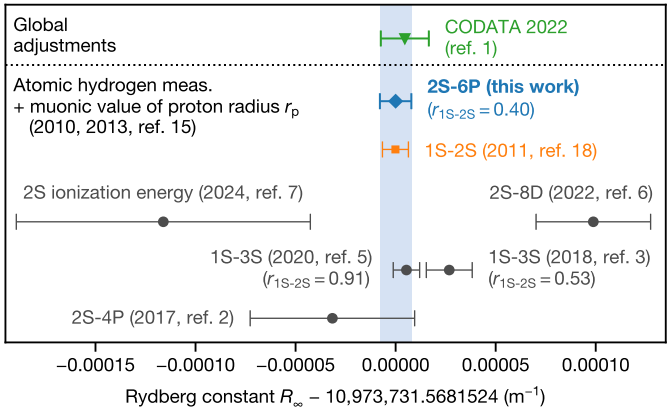

**Extended Data Fig. 2 | Rydberg constant $R_\infty$ from atomic hydrogen.** The $R_\infty$ values are determined by combining each atomic hydrogen measurement[2,3,5–7,18] with the proton radius $r_p$ from muonic hydrogen spectroscopy[8,9,15] and equation (1). The large, correlated QED uncertainty of equation (1) for S-states (along with the common $r_p$ input) leads to a large correlation between some $R_\infty$ values, reducing their sensitivity as QED tests ($r_{1S-2S}$ is the Pearson correlation coefficient with $R_\infty$ from 1S–2S measurement[18]; $r_{1S-2S} < 0.1$ if not shown). In particular, $r_{1S-2S} = 0.91$ for the 2020 1S–3S measurement because of the dominant 1S level QED uncertainty[5], whereas $r_{1S-2S} = 0.40$ for the 2S–6P measurement because of the (eightfold) lower QED uncertainty of the 2S level. For reference, we show the value from the CODATA 2022 global adjustment[1], which includes an uncertainty expansion factor to account for the scatter of the data. The addition of the high-precision, relatively uncorrelated 2S–6P measurement is likely to greatly reduce the effective scatter of future global adjustments, as less precise measurements are removed from the adjustment. Error bars show one-standard-deviation uncertainties.

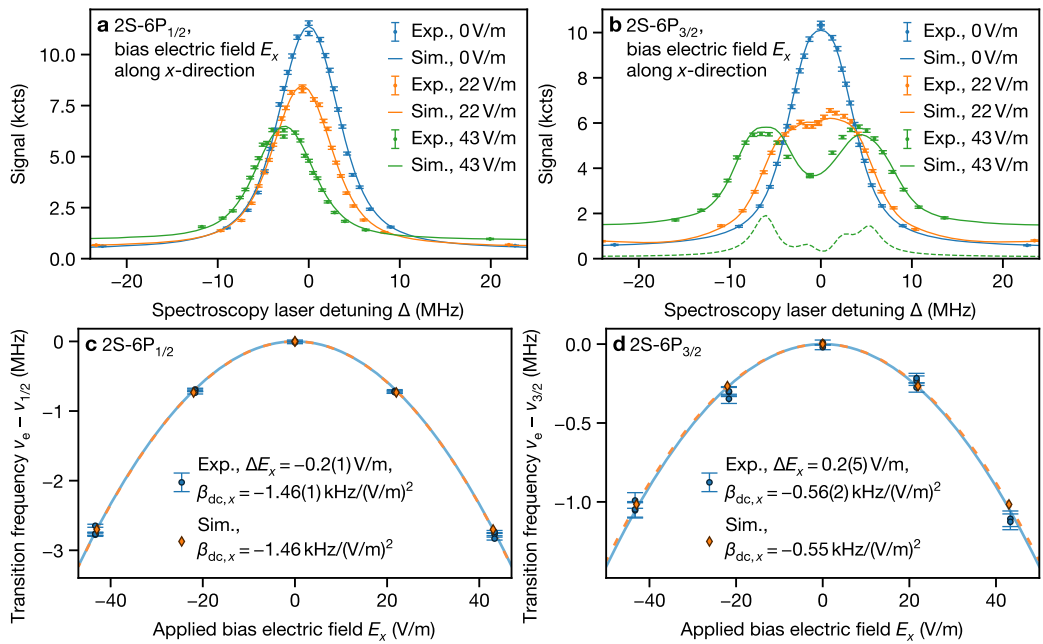

**Extended Data Fig. 3 | In situ determination of stray electric fields. a**, The line of the 2S–6P$_{1/2}$ transition (circles) shifts as a bias electric field $E_x$ (here along the $x$-direction; Fig. 2b) is applied. Velocity group $\tau_9$ is shown, as its mean speed is closest to the mean speed $\langle \bar{v} \rangle$ of all velocity groups. Solid lines show the simulated line shape for each value of $E_x$ (scaled to match experimental data). A Voigt line shape fit (not shown) is used to find the resonance frequency. **b**, The line shape of the 2S–6P$_{3/2}$ transition (circles) splits into two distinct components as $E_x$ is increased and excitations of 6S and 6D levels become allowed. The simulated line shapes (solid lines) reproduce this splitting well and describe the relative amplitudes of the resulting components reasonably well. A simulation of a single atomic trajectory without Doppler broadening (dashed green line; scaled for visibility; $E_x = 43$ V m$^{-1}$) reveals the more complex underlying substructure. A Voigt doublet line shape fit (not shown) is used to find the centre-of-mass resonance frequency of the line shape. **c**, A single stray field measurement using the 2S–6P$_{1/2}$ transition. The Doppler-free transition frequency $\nu_e(E_x)$ (blue circles; error bars show statistical uncertainty only), found by the usual Doppler shift extrapolation, varies quadratically with $E_x$. A quadratic fit (solid blue line) gives the stray electric field component $\Delta E_x$ (Extended Data Fig. 4) and quadratic dc-Stark shift coefficient $\beta_{dc}$. The simulation (orange diamonds) and its fit (dashed orange line) agree with the experiment. The field-induced shift of the transition frequencies coincides with the energy shift of the excited level of the 2S–6P$_{1/2}$ transition. **d**, Similar to panel **c** but using the 2S–6P$_{3/2}$ transition, again showing good agreement between experiment and simulation. The shift of the centre-of-mass transition frequency is much smaller than both the shift of either the resolved line shape components of panel **b** or the energy shift of the excited level of the 2S–6P$_{3/2}$ transition. Only results for the top detector are shown, as the detector signals are highly correlated. See Methods and Supplementary Methods for details.

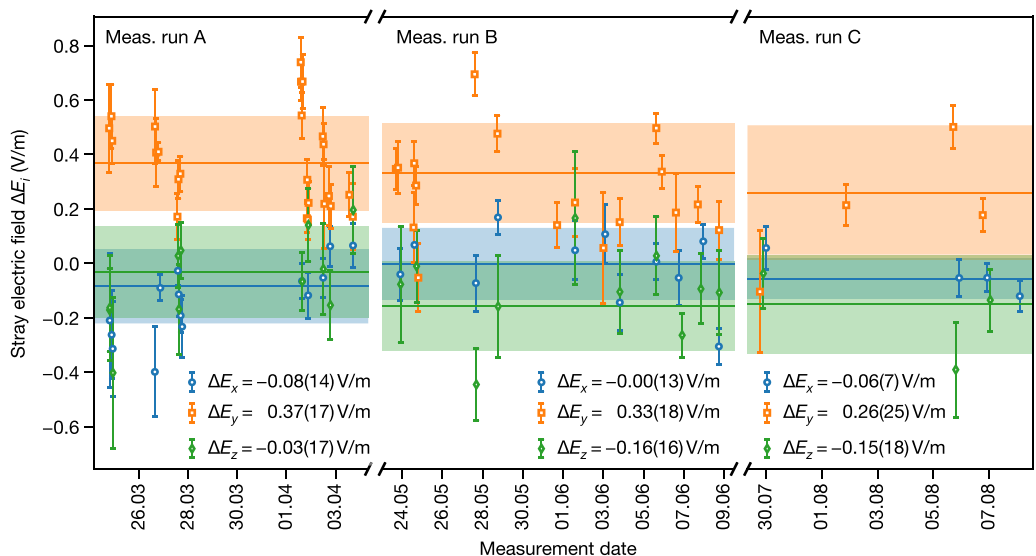

**Extended Data Fig. 4 | Stray electric fields during measurement runs.** The stray electric field components $\Delta E_x$ (blue circles), $\Delta E_y$ (orange squares) and $\Delta E_z$ (green diamonds) in the 2S–6P spectroscopy region, as determined by spectroscopy of the 2S–6P$_{1/2}$ transition with applied bias electric fields (Extended Data Fig. 3 and Methods). See Fig. 2b for the coordinate system definition. Error bars show statistical uncertainties of the detector-averaged data. Shaded areas show the standard deviation for each component and measurement run, with weighted means and standard deviations (in parentheses) shown in the legends. The stray electric field predominantly points along the $y$-direction, which is the axis of the detector cylinder.

# Article

**Extended Data Table 1 | Contributions to the SM prediction $\nu_{2S-6P,SM}$ of the 2S–6P fine-structure centroid**

| Contribution | Value (Hz) | QED-only uncertainty (Hz) |
|---|---:|---:|
| Dirac eigenvalue ($cR_\infty f^{Dirac}$) | 730,691,293,379,477.0 | — |
| Lamb shift $\mathcal{L}$(2S-6P) | −1,044,768,691.8 | 178.6 |
|   Relativistic recoil | −340,939.5 | 0.1 |
|   One-photon corrections ($\propto \alpha^2 \times \alpha^3$) | | |
|     Self-energy | −1,071,052,910.8 | 0.8 |
|     Vacuum polarization | 26,853,096.2 | 0.0 |
|     Muonic ($\mu^+\mu^-$) vacuum pol. | 633.5 | 0.0 |
|     Hadronic vacuum pol. | 425.1 | 10.2 |
|   Two-photon corrections ($\propto \alpha^2 \times \alpha^4$) | −91,497.4 | 126.4 |
|   Three-photon corrections ($\propto \alpha^2 \times \alpha^5$) | −214.6 | 45.8 |
|   Finite nuclear size and polarizability | | |
|     $cR_\infty C_{NS} r_p^2 \propto \alpha^2 \times \alpha^2$ | −138,130.9 | 0.0 |
|     $\propto \alpha^2 \times \alpha^3$ | 13.6 | 1.7 |
|     $\propto \alpha^2 \times \alpha^4$ | −123.5 | 52.8 |
|     $\propto \alpha^2 \times \alpha^5$ | 0.5 | 0.2 |
|   Radiative-recoil corrections | 1540.2 | 102.2 |
|   Nucleus self-energy | −584.2 | 21.6 |
| Total | 730,690,248,610,785.2 | 178.6 |

The naming and grouping of the contributions follow ref. 1. Indented entries detail subcontributions to the Lamb shift (that is, the sum of all QED corrections). The values are calculated by combining equation (1) with the measurement of the 1S–2S transition frequency[18] and the muonic value of $r_p$ (ref. 15). $\alpha^2 \times \alpha^n$ indicates the leading order in $\alpha$ of selected contributions, with the first $\alpha^2$ factor absorbed in $R_\infty$ in equation (1). The QED-only uncertainty is the uncertainty excluding contributions from $r_p$, $\alpha$, $m_p/m_e$ and $\nu_{1S-2S}$. See Methods for details.

**Extended Data Table 2 | Definition and properties of atomic velocity groups $\tau_i$**

| Velocity group | $\tau$ (µs) | $P_{2S}$ | $\bar{v}$ (m/s) | $\Delta v_t$ (m/s) | $\Gamma_F$ (MHz) |
|---|---|---|---|---|---|
| $\tau_1$ | 10...60 | $2.1 \times 10^{-2}$ | 253(5) | 3.32(6) | 9.6(4) |
| $\tau_2$ | 60...110 | $2.1 \times 10^{-2}$ | 252(5) | 3.32(6) | 9.6(4) |
| $\tau_3$ | 110...160 | $2.0 \times 10^{-2}$ | 250(5) | 3.29(6) | 9.5(4) |
| $\tau_4$ | 160...210 | $1.9 \times 10^{-2}$ | 247(5) | 3.23(5) | 9.4(4) |
| $\tau_5$ | 210...260 | $1.7 \times 10^{-2}$ | 242(4) | 3.14(5) | 9.2(4) |
| $\tau_6$ | 260...310 | $1.5 \times 10^{-2}$ | 236(4) | 2.99(4) | 8.9(4) |
| $\tau_7$ | 310...360 | $1.3 \times 10^{-2}$ | 227(4) | 2.79(4) | 8.5(4) |
| $\tau_8$ | 360...410 | $1.2 \times 10^{-2}$ | 217(3) | 2.59(3) | 8.1(4) |
| $\tau_9$ | 410...510 | $9.9 \times 10^{-3}$ | 202(3) | 2.32(2) | 7.7(4) |
| $\tau_{10}$ | 510...610 | $8.9 \times 10^{-3}$ | 183(2) | 2.01(2) | 7.1(4) |
| $\tau_{11}$ | 610...710 | $8.6 \times 10^{-3}$ | 165(2) | 1.76(1) | 6.7(3) |
| $\tau_{12}$ | 710...910 | $7.1 \times 10^{-3}$ | 145(2) | 1.50(1) | 6.2(3) |
| $\tau_{13}$ | 910...1210 | $6.4 \times 10^{-3}$ | 120(1) | 1.19(1) | 5.8(3) |
| $\tau_{14}$ | 1210...1510 | $7.0 \times 10^{-3}$ | 98(1) | 0.94(1) | 5.5(4) |
| $\tau_{15}$ | 1510...2010 | $6.0 \times 10^{-3}$ | 81(1) | 0.76(2) | 5.5(5) |
| $\tau_{16}$ | 2010...2560 | $6.4 \times 10^{-3}$ | 65(1) | 0.58(2) | 5.5(6) |

Each velocity group integrates the fluorescence signal over a range of delay times $\tau$, with $\tau=0$ µs set by the blocking of the 1S–2S preparation laser. Some properties are found by numerically modelling the atomic beam and its interaction with the 1S–2S preparation laser and the 2S–6P spectroscopy laser (see Methods and Section 1.1 in the Supplementary Methods): $P_{2S}$ is the average excitation probability to the metastable 2S level; $\bar{v}$ is the mean speed of atoms contributing to the fluorescence signal; $\Delta v_t$ is the FWHM of the transverse velocity distribution of 2S atoms. The modelling has been experimentally verified by comparison with the signal of each velocity group. $\Gamma_F$ is the experimentally observed FWHM linewidth of the 2S–6P transition. All numbers are weighted averages over the data groups given in Extended Data Table 3 (with the weights $w_{2S-6P}$ given therein); numbers in parentheses are weighted standard deviations over the data groups.

# Article

**Extended Data Table 3 | Experimental parameters of data groups**

| Data group | Meas. run | Transition | $\alpha_0$ (mrad) | $\theta_L$ (°) | $P_{\text{2S-6P}}$ (µW) | FCs | Line scans per detector Top | Line scans per detector Bot. | $v_{\text{cut-off}}$ (m/s) | $P_{\text{1S-2S}}$ (W) | $w_{\text{2S-6P}}$ (%) |
|---|---|---|---|---|---|---|---|---|---|---|---|
| G1A | A | 2S-6P$_{1/2}$ | 0.0 | 56.5 | 30 | 13 | 285 | 285 | $31^{+56}_{-33}$ | $1.30^{+0.50}_{-0.40}$ | 5.7 |
| G1B | B | 2S-6P$_{1/2}$ | 0.0 | 56.5 | 30 | 18 | 148 | 141 | $65^{+39}_{-44}$ | $1.15^{+0.15}_{-0.15}$ | 4.6 |
| G1C | C | 2S-6P$_{1/2}$ | 0.0 | 56.5 | 30 | 16 | 77 | 68 | $52^{+32}_{-18}$ | $1.05^{+0.10}_{-0.10}$ | 2.4 |
| G2 | B | 2S-6P$_{1/2}$ | 0.0 | 56.5 | 20 | 18 | 147 | 138 | $65^{+46}_{-41}$ | $1.15^{+0.15}_{-0.15}$ | 3.5 |
| G3 | B | 2S-6P$_{1/2}$ | 0.0 | 56.5 | 10 | 18 | 598 | 564 | $66^{+43}_{-43}$ | $1.15^{+0.15}_{-0.15}$ | 8.8 |
| G4 | B | 2S-6P$_{1/2}$ | 0.0 | 146.5 | 30 | 3 | 34 | 31 | $41^{+26}_{-23}$ | $1.10^{+0.05}_{-0.05}$ | 0.6 |
| G5 | B | 2S-6P$_{1/2}$ | 0.0 | 146.5 | 20 | 3 | 34 | 31 | $42^{+28}_{-27}$ | $1.10^{+0.05}_{0.00}$ | 0.9 |
| G6 | B | 2S-6P$_{1/2}$ | 0.0 | 146.5 | 10 | 3 | 132 | 119 | $42^{+34}_{-29}$ | $1.10^{+0.05}_{-0.05}$ | 3.0 |
| G7A | A | 2S-6P$_{3/2}$ | 0.0 | 56.5 | 15 | 3 | 162 | 162 | $32^{+34}_{-28}$ | $1.25^{+0.10}_{-0.30}$ | 8.3 |
| G7B | B | 2S-6P$_{3/2}$ | 0.0 | 56.5 | 15 | 17 | 143 | 116 | $57^{+42}_{-43}$ | $1.05^{+0.15}_{-0.10}$ | 13.4 |
| G8 | B | 2S-6P$_{3/2}$ | 0.0 | 56.5 | 10 | 18 | 151 | 124 | $56^{+39}_{-44}$ | $1.05^{+0.15}_{-0.10}$ | 9.2 |
| G9 | B | 2S-6P$_{3/2}$ | 0.0 | 56.5 | 5 | 18 | 568 | 461 | $57^{+43}_{-49}$ | $1.05^{+0.15}_{-0.10}$ | 25.7 |
| G10 | B | 2S-6P$_{3/2}$ | 0.0 | 146.5 | 15 | 2 | 21 | 20 | $42^{+23}_{-19}$ | $1.10^{+0.05}_{-0.05}$ | 2.4 |
| G11 | B | 2S-6P$_{3/2}$ | 0.0 | 146.5 | 10 | 2 | 22 | 21 | $41^{+26}_{-19}$ | $1.10^{+0.05}_{-0.05}$ | 1.7 |
| G12 | B | 2S-6P$_{3/2}$ | 0.0 | 146.5 | 5 | 2 | 87 | 80 | $36^{+35}_{-23}$ | $1.10^{+0.05}_{-0.05}$ | 5.9 |
| G13 | C | 2S-6P$_{1/2}$ | ±7.5 | 56.5 | 30 | 3 | 106 | 106 | $61^{+33}_{-22}$ | $1.10^{+0.05}_{-0.05}$ | 0.8 |
| G14 | C | 2S-6P$_{1/2}$ | ±12.0 | 56.5 | 30 | 11 | 440 | 394 | $63^{+38}_{-43}$ | $1.05^{+0.10}_{-0.15}$ | 2.9 |

Experimental data taken in the same measurement run, and for the same transition, atomic beam offset angle $\alpha_0$, laser polarization angle $\theta_L$ and spectroscopy laser power $P_{\text{2S-6P}}$ are grouped, forming the 17 data groups listed here. Mean values of cut-off speed $v_{\text{cut-off}}$ and preparation laser power $P_{\text{1S-2S}}$ are given, along with the differences to the maximum (superscript) and minimum (subscript) values. Note that approximate values of $v_{\text{cut-off}}$ were used to determine simulation corrections (Extended Data Table 5). FCs, number of freezing cycles contained in the group; line scans per detector, number of line scans contained in the group for top and bottom (bot.) detectors; $w_{\text{2S-6P}}$, statistical weight in the 2S–6P fine-structure centroid $v_{\text{2S-6P}}$. See Methods for details.

**Extended Data Table 4 | Corrections Δν and uncertainties σ for the determination of the 2S–6P$_{1/2}$ ($\nu_{1/2}$) and the 2S–6P$_{3/2}$ ($\nu_{3/2}$) transition frequencies and the 6P fine-structure splitting Δν$_{FS}$(6P)**

| Contribution | $\nu_{1/2}$ Δν (kHz) | $\nu_{1/2}$ σ (kHz) | $\nu_{3/2}$ Δν (kHz) | $\nu_{3/2}$ σ (kHz) | Δν$_{FS}$(6P) Δν (kHz) | Δν$_{FS}$(6P) σ (kHz) |
|---|---|---|---|---|---|---|
| First-order Doppler shift | −0.32 | 0.49 | 0.67 | 0.60 | 1.00 | 0.77 |
|   Extrapolation (statistical) | −0.32 | 0.49 | 0.67 | 0.60 | 1.00 | 0.77 |
|   Simulation of atom speeds | — | 0.02 | — | 0.01 | — | 0.02 |
| Simulation corrections | 0.88 | 0.43 | 1.14 | 0.22 | 0.26 | 0.53 |
|   Light force shift | 0.78 | 0.23 | 1.33 | 0.16 | 0.54 | 0.14 |
|   Quantum interference shift | 0.25 | 0.36 | −0.05 | 0.15 | −0.29 | 0.51 |
|   Second-order Doppler shift | −0.15 | 0.01 | −0.14 | 0.01 | 0.00 | 0.00 |
| dc-Stark shift | 0.20 | 0.21 | −0.02 | 0.06 | −0.22 | 0.23 |
| BBR-induced shift | 0.28 | 0.01 | 0.28 | 0.01 | 0.00 | 0.00 |
| Zeeman shift | 0.00 | 0.02 | 0.00 | 0.11 | 0.00 | 0.09 |
| Pressure shift | 0.00 | 0.02 | 0.00 | 0.02 | 0.00 | 0.00 |
| Sampling bias | 0.00 | 0.04 | 0.00 | 0.08 | 0.00 | 0.04 |
| Signal background | 0.00 | 0.03 | 0.00 | 0.04 | 0.00 | 0.05 |
| Laser spectrum | 0.00 | 0.07 | 0.00 | 0.07 | 0.00 | 0.00 |
| Frequency standard | 0.02 | 0.01 | 0.01 | 0.01 | 0.00 | 0.00 |
| Subtotal (experiment-specific contributions) | 1.06 | 0.69 | 2.08 | 0.66 | 1.04 | 0.97 |
| Recoil shift | −1176.03 | 0.00 | −1176.03 | 0.00 | 0.00 | 0.00 |
| Total (all corrections) | −1174.97 | 0.69 | −1173.95 | 0.66 | 1.04 | 0.97 |

All uncertainties correspond to one standard deviation. Indented entries detail subcontributions to the first-order Doppler shift and simulation corrections. The sum of the subcontributions may differ from the given total owing to rounding. BBR, blackbody radiation.

**Extended Data Table 5 | Contributions to simulation correction uncertainties for the 2S–6P$_{1/2}$ ($\nu_{1/2}$) and 2S–6P$_{3/2}$ ($\nu_{3/2}$) transition frequencies and the 6P fine-structure centroid $\nu_{2S-6P}$ and splitting $\Delta\nu_{FS}(6P)$**

| Contribution | Variation | Sim. | $\nu_{1/2}$ $\sigma$ (Hz) | $\nu_{3/2}$ $\sigma$ (Hz) | $\nu_{2S-6P}$ $\sigma$ (Hz) | $r$ | $\Delta\nu_{FS}(6P)$ $\sigma$ (Hz) | $r$ |
|---|---|---|---|---|---|---|---|---|
| Nozzle temperature $T_N$ | ±100 mK | Total | 84 | 25 | 33 | 0.01 | 77 | 0.42 |
| | | LFS | 80 | 28 | 34 | 0.08 | 74 | 0.38 |
| | | QI | 3 | 1 | 1 | −0.18 | 3 | 0.26 |
| | | SOD | 1 | 1 | 1 | 0.78 | 0 | 0.97 |
| Nozzle radius $r_1$ | $^{+0.1}_{-0.5}$ mm* | Total | 89 | 34 | 52 | 1.00 | 55 | 1.00 |
| | | LFS | 43 | 42 | 42 | 1.00 | 5 | 0.99 |
| | | QI | 34 | 19 | 2 | −0.99 | 53 | −1.00 |
| | | SOD | 11 | 11 | 11 | 1.00 | 0 | 1.00 |
| Cut-off speed $v_{cut-off}$ | min.–max.*,† | Total | 21 | 23 | 16 | −0.19 | 31 | −0.05 |
| | | LFS | 23 | 27 | 26 | 1.00 | 16 | 0.80 |
| | | QI | 16 | 6 | 3 | −0.79 | 19 | −0.40 |
| | | SOD | 7 | 7 | 7 | 1.00 | 0 | 1.00 |
| Atomic beam aperture width | ±0.1 mm | Total | 133 | 32 | 66 | 1.00 | 100 | 1.00 |
| | | LFS | 138 | 40 | 72 | 1.00 | 98 | 1.00 |
| | | QI | 4 | 7 | 6 | 1.00 | 7 | 0.29 |
| | | SOD | 1 | 0 | 0 | −1.00 | 1 | −1.00 |
| Atomic beam offset angle $\alpha_0$ | ±1 mrad | Total | 139 | 126 | 123 | 0.74 | 41 | 0.96 |
| | | LFS | 137 | 130 | 123 | 0.72 | 35 | 0.97 |
| | | QI | 4 | 3 | 1 | −0.83 | 7 | −1.00 |
| | | SOD | 0 | 0 | 0 | 0.90 | 0 | 0.78 |
| 1S-2S preparation laser power $P_{1S-2S}$ and detuning | min.–max.*,‡ | Total | 67 | 33 | 45 | 1.00 | 33 | 1.00 |
| | | LFS | 54 | 34 | 41 | 1.00 | 21 | 1.00 |
| | | QI | 8 | 4 | 4 | 0.07 | 12 | −1.00 |
| | | SOD | 5 | 4 | 4 | 0.84 | 2 | 0.89 |
| 2S-6P spectroscopy laser power $P_{2S-6P}$ | ±10 %* | Total | 59 | 53 | 55 | 1.00 | 7 | 1.00 |
| | | LFS | 35 | 62 | 53 | 1.00 | 28 | 1.00 |
| | | QI | 25 | 9 | 2 | −1.00 | 34 | −1.00 |
| | | SOD | 0 | 0 | 0 | 0.88 | 0 | 0.99 |
| 2S-6P spectroscopy laser polarization angle $\theta_L$ | ±3° | Total | 355 | 152 | 19 | −1.00 | 507 | −1.00 |
| | | LFS | 0 | 0 | 0 | −1.00 | 0 | −1.00 |
| | | QI | 355 | 152 | 19 | −1.00 | 507 | −1.00 |
| | | SOD | 0 | 0 | 0 | −1.00 | 0 | −1.00 |
| Detection efficiency | see § | Total | 33 | 14 | 2 | −0.99 | 47 | −1.00 |
| | | LFS | 0 | 0 | 0 | 0.33 | 0 | 0.33 |
| | | QI | 33 | 14 | 2 | −0.99 | 47 | −1.00 |
| | | SOD | 0 | 0 | 0 | −1.00 | 0 | −1.00 |
| Monte Carlo uncertainty of trajectory simulation | 10 rep.‖ | Total | 28 | 16 | 14 | 0.00 | 32 | 0.00 |
| | | LFS | 28 | 16 | 14 | 0.00 | 32 | 0.00 |
| | | QI | 2 | 2 | 1 | 0.00 | 2 | 0.00 |
| | | SOD | 0 | 0 | 0 | 0.00 | 0 | 0.00 |
| All contributions | | Total | 434 | 216 | 170 | −0.31 | 532 | −0.26 |
| | | LFS | 227 | 164 | 169 | 0.67 | 137 | 0.80 |
| | | QI | 360 | 155 | 20 | −0.99 | 514 | −1.00 |
| | | SOD | 14 | 14 | 14 | 0.98 | 2 | 0.98 |

For each contribution, the simulations were repeated with corresponding input parameters adjusted by the given variation about their optimal value. The resulting one-standard-deviation uncertainties σ for the light force (LFS), QI and second-order Doppler (SOD) shifts are given, along with their combined uncertainty. Pearson correlation coefficients $r$ between the uncertainties for $\nu_{1/2}$ and $\nu_{3/2}$ corresponding to the determined uncertainties for $\nu_{2S-6P}$ and $\Delta\nu_{FS}(6P)$ are also given. The simulation correction uncertainties given in Table 1 and Extended Data Table 4 correspond to the uncertainties given here added in quadrature over the contributions, as reproduced in the last row. The effect on the mean speeds of the velocity groups and thereby the transition frequencies is excluded here and listed separately in Table 1 and Extended Data Table 4. See Section 1.1 in the Supplementary Methods for details. *A multiplicative factor of $1/\sqrt{3}$ is applied to convert the half-width of box-like variations to one-standard-deviation uncertainties. †Mean, minimum and maximum values of $v_{cut-off}$ were taken to be 30, −10 and 90 m s$^{-1}$ for measurement run A and 65, 0 and 130 m s$^{-1}$ for measurement runs B and C (see Extended Data Table 3 for the determined values of $v_{cut-off}$ for each data group). ‡Mean, minimum and maximum values of $P_{1S-2S}$ used for each data group are listed in Extended Data Table 3. The detuning for each value of $P_{1S-2S}$ was found by a linear regression of measured detunings as a function of $P_{1S-2S}$ in each data group. §Detection efficiency uncertainty is estimated by varying the assumed transparencies of the meshes inside the detector cylinder and repeating the Monte Carlo particle tracing simulation of the spatial detection efficiency (see Section 1.2 in the Supplementary Methods). ‖The standard deviation over ten repetitions with randomly drawn trajectories is used as an estimate of the uncertainty from the randomness of the Monte Carlo trajectory simulation.

## Supplementary information

# Sub-part-per-trillion test of the Standard Model with atomic hydrogen

Supplementary Methods for

*Sub-part-per-trillion test of the Standard Model with atomic hydrogen*

Lothar Maisenbacher[1,4*], Vitaly Wirthl[1], Arthur Matveev[1],
Alexey Grinin[1,5], Randolf Pohl[2], Theodor W. Hänsch[1,3], Thomas Udem[1,3]

[1]Max-Planck-Institut für Quantenoptik, Garching, Germany.
[2]Johannes Gutenberg-Universität Mainz, Mainz, Germany.
[3]Ludwig-Maximilians-Universität München, München, Germany.
[4]Present address: University of California, Berkeley, Berkeley, CA, USA.
[5]Present address: Northwestern University, Evanston, IL, USA.

*Corresponding author. E-mail: lothar.maisenbacher@mpq.mpg.de

# 1 Modeling

## 1.1 Modeling of atomic beam and fluorescence line shape

This section summarizes the modeling of the atomic beam and fluorescence line shape of the 2S-6P transition, which is described in detail in Sections 5.2 and 5.3 of [1], with some minor modifications introduced here.

First, the initial positions, directions, and speeds $v$ of atomic trajectories originating from the nozzle are randomly sampled from, respectively, a uniform distribution over the nozzle orifice, a cosine law distribution, and the modified Maxwell-Boltzmann flux distribution for a given cut-off speed $v_{\text{cut-off}}$ (see main text). Trajectories not reaching the 2S-6P spectroscopy region are rejected, and the process is repeated until a set of $N_{\text{traj}}$ is found ($N_{\text{traj}} = 1 \times 10^6$ and $4 \times 10^6$ for data groups G1A–G12 and G13–G14 (see Extended Data Table 3), respectively). These trajectories represent hydrogen atoms in the 1S ground level.

Second, for each of the 16 velocity groups ($i = 1, \ldots, 16$; see Extended Data Table 2) and each trajectory, the excitation probability $P_{\text{2S},i}$ to the metastable 2S level from the interaction with the preparation laser is found by numerically integrating the corresponding optical Bloch equations (OBEs) [2]. The ionization of the 2S level through single-photon absorption from the preparation laser limits the excitation probability. This process is also responsible for the lowered excitation probability for trajectories with low transverse velocity, as is visible in the transverse velocity distribution of the 2S atoms shown on the top of main text Fig. 3a. Importantly, the excitation probability is zero for trajectories that are too fast to both interact with the preparation laser and contribute to a given velocity group. Excluding these trajectories, $P_{\text{2S},i}$ ranges from $P_{\text{2S},1} = 2.1 \times 10^{-2}$ to $P_{\text{2S},16} = 6.4 \times 10^{-3}$ (see Extended Data Table 2).

Third, the excitation from the 2S level to the 6P level by the counter-propagating spectroscopy lasers beams and the subsequent decays are modeled with OBEs. Two different models, the quantum interference (QI) model (see Section 1.2) and light force shift (LFS) model (see Section 1.3), and corresponding sets of OBEs are used in this step. The OBEs of each model include signal equations that contain the expected fluorescence signal of the experimentally detected Lyman decays (split into their three spherical components for the QI model). The solution of the OBEs for each trajectory, velocity group and spectroscopy laser detuning $\Delta$ is found by interpolating precomputed numerical solutions of the OBEs on a three-dimensional, regular grid defined by trajectory speed ($20\,\text{m/s} \ldots 1200\,\text{m/s}$), trajectory transverse angle (QI model; $0.5\,\text{mrad} \ldots 31.5\,\text{mrad}$) or transverse velocity (LFS model; $0\,\text{m/s} \ldots 38\,\text{m/s}$), and spectroscopy laser power ($5\,\mu\text{W} \ldots 50\,\mu\text{W}$ and $1\,\mu\text{W} \ldots 35\,\mu\text{W}$, respectively, for QI and LFS models and 2S-6P$_{1/2}$ transition; twofold lower power for 2S-6P$_{3/2}$ transition). We derive the OBEs using a computer algebra system [3] and numerically integrate them using an 8th-order explicit Runge-Kutta method [4] for the QI model and an implicit Runge-Kutta method (Radau IIA) of variable order [5] for the LFS model. The fluorescence signal for each velocity group, detuning, and decay (split into spherical components for the QI model) is found by summing the corresponding signal equation over all trajectories with the weight of each trajectory given by $P_{\text{2S},i}$. For the QI model, the three spherical components are weighted by their detection efficiency, given by simulated spatial detection efficiency of the two detectors, the component's radiation pattern, and the linear laser polarization angle $\theta_{\text{L}}$, which defines the orientation of the radiation pattern relative to the detector cylinder. The LFS model, on the other hand, does not depend on spatial detection efficiency or $\theta_{\text{L}}$, as it does not distinguish between spherical components. We use the signal from Lyman-$\epsilon$ decays only (summed over their weighted spherical components in case of the QI model) to mimic the experimental signal for each detector, velocity group, and detuning. Including the other Lyman decays, weighted by their detection efficiency, does not significantly influence the simulation corrections (except when static electric fields are included, see Section 1.4).

Fourth, the simulated fluorescence line shape for each detector and velocity group is analyzed in the same way as the experimental line scans (see Methods in the main text), i.e., by fitting either a Voigt or Voigt doublet line shape for data groups G1A–G12 or G13–G14, respectively. To ensure that the (relative) weight of the signal at each detuning $j$ matches that of the experimental data (where the weight $w_j = 1/y_j$ of signal $y_j$ is the inverse square of the expected photon-number shot (Poissonian) noise $\sqrt{y_j}$ of the signal), a constant offset corresponding to the experimentally observed signal background is added to the simulated line shape. The QI shift $\nu_{\text{QI}}$ and the LFS $\nu_{\text{LFS}}$ are given by the resonance frequencies of the line shapes fitted to the QI and LFS simulations, respectively.

The line amplitudes $A$ of the fitted line shapes are used to find the speed distribution of atoms contributing to the fluorescence signal for each velocity group (see Section 5.3.2 of [1]). To this end, the $N_{\text{traj}}$ trajectories are sorted into 200 equal-width speed bins covering the range $v = 0\,\text{m/s} \ldots 1000\,\text{m/s}$, and, for each bin, the simulated line shape is calculated and fitted. This allows us to calculate the line amplitude of each velocity group for an arbitrary cut-off speed $v_{\text{cut-off}}$ by appropriately weighting the line amplitude of each bin. By comparing the simulated and experimental line amplitudes of the velocity groups for each line scan, we find the value of $v_{\text{cut-off}}$ that best describes the experimental data (see Extended Data Table 3). While not strictly necessary, we in practice first determine $v_{\text{cut-off}}$ (using a set of trajectories calculated using an initial value for $v_{\text{cut-off}}$) and then calculate a new set of $N_{\text{traj}}$ trajectories using the optimal $v_{\text{cut-off}}$ for the rest of the analysis. The mean speed $\bar{v}$ and the root-mean-square speed $\bar{v}_{\text{RMS}}$ of each velocity group are found from the weighted mean of the mean speed of the bins, using the line amplitudes as weights. We use the QI model with uniform spatial detection efficiency to determine $v_{\text{cut-off}}$, $\bar{v}$, and $\bar{v}_{\text{RMS}}$; using the LFS model or the QI model with the detectors' spatial detection efficiencies instead does not give significantly different results.

For each experimental data group, the above steps are repeated using the appropriate optimal experimental parameters (see Extended Data Tables 3 and 5). This results in one value each of $\nu_{\text{QI}}$, $\nu_{\text{LFS}}$, $\bar{v}$, and $\bar{v}_{\text{RMS}}$ for each velocity group and each detector (for $\nu_{\text{QI}}$ and $\nu_{\text{LFS}}$) for every data group (see Extended Data Table 2 for the average value of $\bar{v}$ for each velocity group).

The uncertainty of the simulation corrections is found by varying the input parameters within their experimental limits. This is done for each input parameter individually while holding all other input parameters at their optimal values. The input parameters varied, their ranges of variation, and the resulting uncertainties are listed in Extended Data Table 5. When determining these uncertainties, we exclude any effects on $\nu_{\text{e}}$ from the variation in the mean speeds $\bar{v}$ caused by the variation of the input parameters. This allows us to distinguish the effect of a given input parameter variation on the light force, quantum interference, and second-order Doppler shifts from its common-mode effect on the values of $\bar{v}$, which, as explained above, are not sensitive to the specific underlying model. Instead, we include the uncertainty from the variation of $\bar{v}$ separately (see entry "Simulation of atom speeds" in main text Table 1 and Extended Data Table 4). We note that each input parameter variation contributes at most 12 Hz to this uncertainty.

## 1.2 Quantum interference (QI) model

The QI model underlies the QI simulations shown in this work and the estimations of the quantum interference shift given in main text Table 1 and Extended Data Table 4. It is described in detail in Section 2.3.2 of [1] (where it is referred to as "big model"). Briefly, the QI model contains all 148 levels coupled by the 2S→6P excitation and subsequent radiative, electric dipole-allowed (E1) decays. The level energies are taken from [6], which uses the equivalent of Eq. (1) in the main text to tabulate values. The dipole moments are calculated using non-relativistic wave functions as described in [7] (including reduced-mass corrections; relativistic corrections to the dipole moments are of order $\alpha^2$ [8] and can be safely neglected here). All decay rates referenced in this work are derived from this model. Their values are (to 4 significant digits; dcy/s: decays per second): $\Gamma = 3.899\,\text{MHz} = (2\pi \times 3.899)\,\text{Mdcy/s}$, $\Gamma_{\text{e-1S}} = (2\pi \times 3.437)\,\text{Mdcy/s}$, $\Gamma_{\text{det}} = (2\pi \times 3.140)\,\text{Mdcy/s}$ (corresponding to Lyman-$\epsilon$ decay), $\gamma_{\text{e-2S}} = (2\pi \times 462.0)\,\text{kdcy/s}$, and $\gamma_{\text{ei}} = (2\pi \times 153.6)\,\text{kdcy/s}$ or $\gamma_{\text{ei}} = (2\pi \times 306.4)\,\text{kdcy/s}$ for the 2S-6P$_{1/2}$ or 2S-6P$_{3/2}$ transitions, respectively (with direct Balmer-$\delta$ decays contributing $(2\pi \times 151.6)\,\text{kdcy/s}$ or $(2\pi \times 303.3)\,\text{kdcy/s}$ to $\gamma_{\text{ei}}$). A signal equation for each (allowed) spherical component ($\sigma^-$, $\pi$, $\sigma^+$) of each decay between level manifolds with different principal quantum numbers is included, resulting in a total of 42 signal equations. This results in a total of 732 real-valued, nonzero coupled optical Bloch equations. The model includes the two 6P fine-structure manifolds and terms describing cross-damping between their decays (cross-damping terms are included for all decays that share the same principal quantum numbers of both upper and lower levels). It therefore describes line shape distortions from quantum interference between excitation–decay paths that go through either 6P fine-structure manifold but start and end in the same level (the 2S-6P$_{1/2}$ and 2S-6P$_{3/2}$ transitions share the same initial level, but their excited levels belong to the $J = 1/2$ and $J = 3/2$ manifolds, respectively; see main text) [9, 10].

Because the different spherical components of the decays have different radiation patterns, a simulation of the spatial detection efficiency of the top and bottom detectors as a function of the photon emission direction is necessary to predict the fluorescence line shape observed in the experiment. To this end, a Monte Carlo particle tracing simulation of Lyman-$\epsilon$ photons and the photoelectrons ejected by them from the detector cylinder walls is employed, taking into account the different properties of the colloidal-graphite-coated walls and oxidized aluminum walls of the Faraday cage and the detector cylinder, respectively (see Section 4.6.6 of [1]). The uncertainty of the particle tracing simulation is estimated by varying the assumed transparencies of the meshes inside the detector cylinder, as meshes with different transparencies were used in the course of the experiment and as the mesh transparency is affected by the graphite coating, and by repeating the simulation (with $>1 \times 10^8$ photons used in each run) to account for the random nature of the simulation (see Extended Data Table 5). A similar particle tracing simulation was used to simulate the QI shifts in our previous measurement of the 2S-4P transition [11].

The top and bottom detectors are highly symmetric, except in one regard: the top detector has a solid aluminum cap on top of the detector cylinder, while the bottom detector has a graphite-coated mesh at the bottom of the detector cylinder to maintain the vacuum inside the cylinder (see main text Fig. 2b). The resulting difference in spatial detection efficiency leads to approximately 10 % larger QI shifts for the top detector than for the bottom detector.

In the perturbative limit [10], QI shifts are independent of laser power and vanish for a fluorescence detection covering the full ($4\pi$) solid angle. However, this is no longer the case when accounting for saturation and optical pumping, both included in our QI model, and we find two distinct effects to be of particular importance.

First, because of saturation, the ratio between the excitation rates of the off-resonantly coupled and the near-resonantly coupled 6P fine-structure components will increase with spectroscopy laser power. That is, the perturbing transition increases in strength relative to the perturbed transition, which in turn increases the QI shift.

Second, the back decay to the initial 2S level from the excited 6P levels introduces an asymmetry (with respect to the detuning from the probed transition) in the initial level's population, since the back decay is affected by QI distortions as well. This population asymmetry can then be imprinted on the population of the excited 6P levels by re-excitation and may be detected as signal upon decay. Importantly, the QI shifts from this optical pumping effect are independent of detection geometry. For our laser power range, they are found to be proportional to laser power with slopes of $-20$ Hz/µW and $21$ Hz/µW for the 2S-6P$_{1/2}$ and 2S-6P$_{3/2}$ transitions, respectively.

While the second effect counteracts the saturation effect, overall the maximum QI shifts increase with laser power. For our laser power range and spatial detection efficiency, we find this increase to be approximately linear, amounting to $2.7 \%$/µW and $5.3 \%$/µW of the shifts in the perturbative limit for the 2S-6P$_{1/2}$ and 2S-6P$_{3/2}$ transitions, respectively. Likewise, the two effects lead to a dependence of the magic angle (the polarization angle $\theta_\text{L}$ for which the QI shifts are zero) on laser power, shifting from $\approx 54°$ in the zero-power limit to $\approx 52°$ at the highest laser powers used here.

Besides the QI effect, the presence of both 6P fine-structure manifolds in the model ensures that the dominant contribution to the ac-Stark shift of the 2S-6P transitions is included. It is estimated to be at most (i.e., at the highest spectroscopy laser powers used here) $41$ Hz and $-10$ Hz for the 2S-6P$_{1/2}$ and 2S-6P$_{3/2}$ transitions, respectively. Contributions to the ac-Stark shift from off-resonant excitation of levels with $n \neq 6$, not included in the QI model, are estimated to contribute below $1$ mHz in total for either transition [1].

## 1.3 Light force shift (LFS) model

The LFS model underlies the LFS simulations shown in this work and the LFS corrections given in main text Table 1 and Extended Data Table 4. It is introduced in the main text and described in detail in Section 3.4 of [1]. Briefly, it describes the state of the atoms in the combined basis of a simplified internal energy level scheme and the external momenta along the standing wave formed by the spectroscopy laser beams. The longitudinal motion is treated classically in the form of a time-dependent Rabi frequency in the rest frame of the atom, while the motion perpendicular to both the longitudinal direction and standing wave can be ignored here. The simplified level scheme consists of the initial $2S_{1/2}^{F=0}$, $m_F = 0$ level, either the excited $6P_{1/2}^{F=1}$, $m_F = 0$ or $6P_{3/2}^{F=1}$, $m_F = 0$ level, and a single 1S ground level, along with the (near-)resonant 2S→6P excitation and the 6P→2S and 6P→1S decays. Absorption or stimulated emission of a photon from or into either of the spectroscopy laser beams both changes the atom's internal level occupation and leads to a change in momentum along the beams by $\pm\hbar K_L$.

The effective rate of the 6P→2S back decay is set to $\gamma_{ei}$, which includes the direct and all indirect decays to the $2S_{1/2}^{F=0}$, $m_F = 0$ level. This is a good approximation because the indirect decays only contribute $\approx 1\%$ of $\gamma_{ei}$ (see Tables 2.4 and 2.5 of [1]). All other decays of the 6P level are included in the effective 6P→1S decay rate (given by $\Gamma - \gamma_{ei}$), including those decays that would otherwise lead to 2S levels other than the $2S_{1/2}^{F=0}$, $m_F = 0$ level. This is a good approximation as those 2S levels are not resonantly coupled to the 6P levels and any population reaching them is effectively lost from the system.

The basis is split into two coupled subbases: one containing states before a momentum-changing 6P→2S back decay has taken place, and the other containing states after such a decay has changed the momentum by $\Delta p_{D,1}$ (picked from an appropriate distribution, see below). For the first subbasis, the rates of 6P→2S and 6P→1S decays are set to $\gamma_{ei}$ and $\Gamma - \gamma_{ei}$, respectively, while for the second subbasis, they are set to zero and $\Gamma$. This effectively limits the number of back decays to one, as otherwise the required momentum basis grows with each subsequent back decay. Since $\gamma_{ei}/\Gamma \ll 1$, this is a good approximation, as confirmed by simulations including multiple subsequent back decays [1]. The subbases include coupled states with momenta ranging from $p_0 - 4\hbar K_L$ to $p_0 + 4\hbar K_L$ and $p_1 - 4\hbar K_L$ to $p_1 + 4\hbar K_L$, respectively, where $p_0$ is the initial momentum and $p_1 = p_0 + \Delta p_{D,1}$. The maximum momentum change ($\pm 4\hbar K_L$) that needs to be taken into account for the accuracy required here was found by varying the size of the basis [1]. This results in a total of 27 states (see Fig. 3.2 of [1] for a simplified visualization for $\pm 2\hbar K_L$) and 207 real-valued, nonzero coupled optical Bloch equations (OBEs), including 4 signal equations.

For each set of input parameters, the OBEs are numerically integrated for different values of $\Delta p_{D,1}$, and the resulting simulated fluorescence line shapes are weighted with the distribution of $\Delta p_{D,1}$ along the standing wave (Eq. (74) of [12]; note the different choice of quantization axis) and averaged. We use the Gaussian quadrature rule with 4 points to average over $\Delta p_{D,1}$. The averaged line shapes are fit in the same way as the experimental data to determine their resonance frequency, giving the LFS $\nu_{LFS}$ as shown in main text Fig. 3a, b. The width of the Bragg resonance of Fig. 3b is dominated by power broadening and time-of-flight broadening.

As a test, we use a perturbative analysis of the OBEs [13] at zero $v_x$ and in the limit of small excited state population, which is approximately met here, to find $\nu_{LFS} \approx -|\Omega_0/(2\pi)|^2/(16\Delta\nu_{rec})$, where $\Omega_0$ is the (angular) Rabi frequency of the 2S→6P excitation for each spectroscopy laser beam. Using an effective Rabi frequency of $\Omega_0 = (2\pi \times 109)$ krad/s to approximate the situation of an atom crossing the Gaussian spectroscopy laser beams, each with $P_{2S-6P} = 30\,\mu W$ (15 µW) power for the 2S-6P$_{1/2}$ (2S-6P$_{3/2}$) transition, we find agreement within 60 Hz between this perturbative result and the full simulation.

Finally, we model the experimentally observed fluorescence line shape by summing up the line shapes of a set of atomic trajectories representing the atomic beam (see Section 1.1). This incoherent sum is valid because the incident transverse momentum states, as described by the Wigner function, are only mutually coherent over $2.6 \times 10^{-4}\hbar K_L$ in momentum space (this scale is not enhanced by propagation), which is much smaller than the momentum separation of $2\hbar K_L$ of the coupled states [1].

## 1.4 Modeling of dc-Stark shift

The mechanism underlying the dc-Stark shift is the mixing of the levels of interest with perturbing nearby levels of opposite parity by the electric field $\boldsymbol{E}$. The resulting new eigenstates are not only shifted in energy from their zero-field equivalents, but also no longer parity eigenstates. In particular, excitations from the initial 2S level of the 6S and 6D levels (which now have some admixture of the 6P level) become dipole allowed [1]. We note that the shift of the energy levels scales as $n^7$ [7].

Importantly, we do not directly determine the transition frequency between the levels of interest but instead measure and fit a fluorescence line shape to find the transition frequency. If excitations to perturbing levels contribute to this line shape, the dc-Stark shift $\Delta\nu_{\text{dc}}$ of the transition frequency determined in this way is no longer adequately described by the energy shift of the levels of interest, but instead must be found from a fit to the line shape at field $\boldsymbol{E}$. Indeed, it can be shown within second-order perturbation theory (see Section 2.4 of [1]) that the dc-Stark shift of the center of mass of the mixed levels vanishes. However, this description is only adequate when the energy separations of the levels are much larger than their linewidths, which is not always the case here. In addition, a line shape fit is not necessarily equivalent to finding the center of mass of the mixed levels. Nevertheless, we may expect some cancellation and indeed observe this for the 2S-6P$_{3/2}$ transition (see below).

To investigate this, we simulate corresponding line shapes by extending our QI model (see Section 1.2) to include mixing terms from a static electric field with strength $E = |\boldsymbol{E}|$, either parallel or perpendicular to the quantization axis (set by the linear spectroscopy laser polarization), the mixed 6S and 6D levels, and their decays along with the necessary additional intermediate levels. The simulation also includes the field-induced mixing of the 2S levels with the 2P levels, but the resulting shift of the initial 2S level (which we find to be $0.4\,\text{Hz}/(\text{V/m})^2$) is negligible here. Because we treat the electric field as a perturbation, the relevant 6P levels (for which $F = 1$) only mix with levels with $F \neq 1$ for the case of electric field parallel to the quantization axis, and only with levels with $F \neq 0$ for the perpendicular case. This results in a total of 2070 and 5880 real-valued, nonzero coupled optical Bloch equations (OBEs) for the parallel and perpendicular cases, respectively. Due to the resulting computational cost, we only calculate single atomic trajectories, adding the experimentally observed Doppler broadening by convolving the simulated line shape with a Gaussian, except for the case of the 2S-6P$_{3/2}$ transition and electric field perpendicular to the quantization axis (see below), where we average over a set of trajectories representing the atomic beam. As for the QI simulation, the three spherical components of each decay are weighted by their detection efficiency and summed up. Since the laser polarization lies in the $y$-$z$-plane in the experiment (see main text Fig. 2b), an electric field along the $x$-direction is purely perpendicular to the quantization axis (i.e., $\beta_{\text{dc},x} \equiv \beta_{\text{dc},\perp}$), while electric fields along the $y$- and $z$-directions can be decomposed into components parallel and perpendicular to the quantization axis (i.e., $\beta_{\text{dc},y} = \cos^2(\theta_{\text{L}})\beta_{\text{dc},\perp} + \sin^2(\theta_{\text{L}})\beta_{\text{dc},\parallel}$ and $\beta_{\text{dc},z} = \sin^2(\theta_{\text{L}})\beta_{\text{dc},\perp} + \cos^2(\theta_{\text{L}})\beta_{\text{dc},\parallel}$ ). Note that while this decomposition is an approximation because the line shapes are different for parallel and perpendicular fields, we find reasonable or better agreement with the experimental data (see below). The simulations are performed for both the stray-field regime ($E < 1\,\text{V/m}$) and the bias-field regime ($E = 10\,\text{V/m} \ldots 45\,\text{V/m}$) as defined in the Methods in the main text.

For the 2S-6P$_{1/2}$ transition, the perturbation to its excited 6P$_{1/2}^{F=1}$, $m_F = 0$ level is from mixing with the 6S$_{1/2}^{F=0}$ and 6S$_{1/2}^{F=1}$ levels, which are higher in energy by 34 MHz and 41 MHz, respectively, and, to a lesser degree, the 6D$_{3/2}^{F=1}$ and 6D$_{3/2}^{F=2}$ levels, which are higher in energy by 405 MHz, approximately the 6P fine-structure splitting [6]. All perturbing levels are therefore well outside the natural and experimental linewidth and we expect the dc-Stark shift of the 2S-6P$_{1/2}$ transition to be given by the shift of the 6P$_{1/2}^{F=1}$, $m_F = 0$ level (and the much smaller shift of the initial 2S level). Likewise, while the line of the transition will shift, its line shape will remain well-described by a Voigt line shape, as observed in the experiment (see Extended Data Fig. 3a). From second-order perturbation theory for the shift of the 6P$_{1/2}^{F=1}$, $m_F = 0$ level, we obtain $\beta_{\text{dc},\parallel} = -1.75\,\text{kHz}/(\text{V/m})^2$ and $\beta_{\text{dc},\perp} = -1.51\,\text{kHz}/(\text{V/m})^2$ for the static electric field parallel and perpendicular to the quantization axis, respectively. The line shape simulations, which reproduce the experimental line shape well (see solid lines in Extended Data Fig. 3a), give values for $\beta_{\text{dc},\parallel}$ and $\beta_{\text{dc},\perp}$ that agree within 4 % (where spectroscopy laser powers and atom speed and transverse velocity were varied) with the perturbative values for both the stray- and bias-field regimes. Experimentally (for $\theta_\text{L} = 56.5°$), we find $\beta_{\text{dc},x} = -1.469(4)\,\text{kHz}/(\text{V}/\text{m})^2$, $\beta_{\text{dc},y} = -1.530(3)\,\text{kHz}/(\text{V}/\text{m})^2$, and $\beta_{\text{dc},z} = -1.707(7)\,\text{kHz}/(\text{V}/\text{m})^2$ (statistical uncertainties only), which agree within 4 % with the bias-field regime simulations (see Extended Data Fig. 3c for a measurement of $\beta_{\text{dc},x}$ along with the simulation). We choose to use the experimental values of $\beta_\text{dc}$ to determine the dc-Stark shift from stray electric fields for the 2S-6P$_{1/2}$ transition, expanding their uncertainty to cover both stray- and bias-field regimes and theory and simulation results ($\beta_{\text{dc},x} = -1.47(4)\,\text{kHz}/(\text{V}/\text{m})^2$, $\beta_{\text{dc},y} = -1.53(5)\,\text{kHz}/(\text{V}/\text{m})^2$, and $\beta_{\text{dc},z} = -1.71(5)\,\text{kHz}/(\text{V}/\text{m})^2$).

For the 2S-6P$_{3/2}$ transition the situation is quite different. This is because the perturbation of its excited 6P$_{3/2}^{F=1}$, $m_F = 0$ level is dominated by the 6D$_{3/2}^{F=2}$ and 6D$_{3/2}^{F=1}$ levels, which are only separated in energy by 57 kHz and $-469$ kHz, respectively [6]. They are therefore well within the natural linewidth of the 2S-6P$_{3/2}$ transition, and the observed dc-Stark shift of the transition frequency determined from the line shape is not expected to correspond to the shift of the 6P$_{3/2}^{F=1}$, $m_F = 0$ level. Because of the strong mixing resulting from the small energy separation, we expect the line shape to be strongly perturbed, and indeed we experimentally observe a splitting of the line shape into two distinct components (see Extended Data Fig. 3b). We use a Voigt doublet line shape to fit the experimental and simulated line shape. The other perturbing levels, 6S$_{1/2}^{F=0}$ and 6S$_{1/2}^{F=1}$ (lower in energy by 365 MHz and 371 MHz, respectively), and 6D$_{5/2}^{F=2}$ (higher in energy by 135 MHz), are again well-separated compared to the linewidth and mix much less strongly with the 6P$_{3/2}^{F=1}$, $m_F = 0$ level.

We first consider the case of the electric field perpendicular to the quantization axis, for which both the 6D$_{3/2}^{F=2}$ and the 6D$_{3/2}^{F=1}$ levels are mixed within the linewidth. Our simulation (solid lines in Extended Data Fig. 3b) reproduces the splitting of the experimental line shape well and describes the resulting components' relative amplitudes reasonably well within $\approx 20\,\%$. The simulation of a single trajectory without Doppler broadening and at a high bias field (green dashed line in Extended Data Fig. 3b) reveals a more complex substructure. In the bias-field regime, the simulations nevertheless result in a quadratic dc-Stark shift with $\beta_{\text{dc},\perp} = -0.55(3)\,\text{kHz}/(\text{V}/\text{m})^2$ (orange diamonds and dashed line in Extended Data Fig. 3d). This agrees within 5 % with the experimental value of $\beta_{\text{dc},x} = -0.573(7)\,\text{kHz}/(\text{V}/\text{m})^2$ (statistical uncertainty only; see blue circles and solid line in Extended Data Fig. 3d for a single measurement). Both values agree within 15 % with our second-order perturbation theory result of $\beta_{\text{dc},\perp} = -0.50\,\text{kHz}/(\text{V}/\text{m})^2$ for the shift of the 6P$_{3/2}^{F=1}$, $m_F = 0$ level when the mixing with the 6D$_{3/2}^{F=2}$ and 6D$_{3/2}^{F=1}$ levels is ignored, i.e., assuming that there is no net contribution from levels within the linewidth. This can be understood as the line shape fit approximately finding the center of mass of the (split) line shape and thereby cancelling the shift from the perturbing levels causing the splitting. We use $\beta_{\text{dc},\perp} = -0.55(3)\,\text{kHz}/(\text{V}/\text{m})^2$ as our estimate in the bias-field regime.

In the stray-field regime, where the splitting is not resolved, the simulations reveal a more complex behavior. There is a substantial dependence of $\beta_{\text{dc},\perp}$ on the experimental parameters of each trajectory, especially on the transverse velocity, which we attribute to the interplay of line splitting due to the electric field and due to the Doppler shift. This is addressed by averaging over multiple trajectories representing the atomic beam, as explained above. Most strikingly, $\beta_{\text{dc},\perp}$ strongly depends on which Lyman decay is observed (with the dc-Stark shift's quadratic behavior maintained for each decay within the stray-field regime), e.g., we find $1.39\,\text{kHz}/(\text{V}/\text{m})^2$ for Lyman-$\epsilon$ and $-17.7\,\text{kHz}/(\text{V}/\text{m})^2$ for Lyman-$\alpha$ decays (at $P_{\text{2S-6P}} = 15\,\mu\text{W}$), which approximately constitute 97 % and 1.5 % of the fluorescence signal, respectively. We address this by summing up the Lyman decays with their estimated detection efficiency (see Fig. B1 of [1]). Overall, we estimate $\beta_{\text{dc},\perp}$ in the stray-field regime as $0.56(75)\,\text{kHz}/(\text{V}/\text{m})^2$ (note the opposite sign to $\beta_{\text{dc},\perp}$ in the bias-field regime).

Next, we consider the simpler case of the electric field parallel to the quantization axis, for which only the $6D_{3/2}^{F=2}$ level is mixed within the linewidth. The agreement between the simulated and experimental line shapes is similar to the perpendicular case. However, as opposed to the perpendicular case, the simulated values of $\beta_{\text{dc},\perp}$ show no strong dependence on the experimental parameters or on which Lyman decay is detected. Importantly, there is no drastically different behavior for the stray- and bias-field regimes, which we attribute to the presence of only one mixed level within the linewidth. Overall, we find $\beta_{\text{dc},\parallel} = -0.48(5)\,\text{kHz}/(\text{V/m})^2$. Remarkably, this simulated value is much smaller than the second-order perturbation theory result of $\beta_{\text{dc},\parallel} = -33.6\,\text{kHz}/(\text{V/m})^2$ for the shift of the $6P_{3/2}^{F=1}$, $m_F=0$ level when all perturbing levels are included, but in excellent agreement with the result ($\beta_{\text{dc},\parallel} = -0.48\,\text{kHz}/(\text{V/m})^2$) when the mixing with the $6D_{3/2}^{F=2}$ level is ignored, again demonstrating the strong cancellation of the shift when finding the center of mass of the split line.

Using the bias-field regime value of $\beta_{\text{dc},\perp}$ from above, we find the simulated values of $\beta_{\text{dc},y}$ and $\beta_{\text{dc},z}$ in the bias-field regime (for $\theta_{\text{L}} = 56.5°$) to be $-0.53(3)\,\text{kHz}/(\text{V/m})^2$ and $-0.50(4)\,\text{kHz}/(\text{V/m})^2$, respectively, showing good agreement with the experimental value of $\beta_{\text{dc},y}$ ($-0.540(5)\,\text{kHz}/(\text{V/m})^2$) and reasonable agreement for $\beta_{\text{dc},z}$ ($-0.422(18)\,\text{kHz}/(\text{V/m})^2$). Likewise, we find the simulated values of $\beta_{\text{dc},y}$ and $\beta_{\text{dc},z}$ in the stray-field regime to be $0.24(52)\,\text{kHz}/(\text{V/m})^2$ and $-0.16(24)\,\text{kHz}/(\text{V/m})^2$, respectively. These two values, along with the stray-field-regime value of $\beta_{\text{dc},x} \equiv \beta_{\text{dc},\perp}$, are used in the determination of the dc-Stark shift of the 2S-$6P_{3/2}$ transition. The Pearson correlation coefficients between $(\beta_{\text{dc},x}, \beta_{\text{dc},y})$, $(\beta_{\text{dc},x}, \beta_{\text{dc},z})$, and $(\beta_{\text{dc},y}, \beta_{\text{dc},z})$ are $r = 0.99$, $r = 0.97$, and $r = 0.95$, respectively.

## 2 Corrections and uncertainties

### 2.1 Second-order Doppler shift

The second-order Doppler shift (SOD) leads to an apparent shift of the atom's transition frequency $\nu$ in the laboratory frame of reference (see Section 2.2.3 of [1]). The average value of the shift for each velocity group with root-mean-square speed $\bar{v}_{\text{RMS}}$ is given by

$$\Delta\nu_{\text{SOD}} = -\frac{\nu}{2}\left(\frac{\bar{v}_{\text{RMS}}}{c}\right)^2. \tag{1}$$

The transition frequency $\nu$ is identical at the required level of accuracy for the two 2S-6P transitions probed here and is known with much lower uncertainty than needed here. $\bar{v}_{\text{RMS}}$ and $\Delta\nu_{\text{SOD}}$ range over $270\,\text{m/s}\ldots 67\,\text{m/s}$ and $-297\,\text{Hz}\ldots -18\,\text{Hz}$, respectively, for the different velocity groups.

Since the magnitude of the (in this geometry always negative) SOD increases (quadratically) with speed, it leads to an apparent negative Doppler slope of $\kappa = -1.7\,\text{Hz}/(\text{m/s})$ in the (linear) Doppler shift extrapolation if not accounted for. Here, we first correct the resonance frequency $\nu_0$ of each velocity group of each line scan by subtracting $\Delta\nu_{\text{SOD}}$, with $\bar{v}_{\text{RMS}}$ determined with our atomic beam simulation, and subsequently performing the Doppler shift extrapolation. The effect of the removed apparent Doppler slope outweighs the shift of the individual velocity groups, and the correction applied to the Doppler-free transition frequencies $\nu_{1/2}$ and $\nu_{3/2}$ amounts to $-0.15(1)\,\text{kHz}$ and $-0.14(1)\,\text{kHz}$, respectively (see Extended Data Table 5 for the uncertainty estimation). The correction is highly correlated ($r = 0.98$) between the 2S-$6P_{1/2}$ and 2S-$6P_{3/2}$ transitions.

### 2.2 Blackbody-radiation (BBR)-induced shift

We calculate the BBR-induced ac-Stark shifts of the 2S and 6P levels following [14] but include fine and hyperfine structure and the Lamb shift for levels with $n \leq 10$. However, because the resulting frequency splittings of the 2S and 6P levels (at most $10\,\text{GHz}$) are far below the peak of the BBR spectrum near room temperature ($\sim 20\,\text{THz}$), this inclusion has a negligible effect at our level of accuracy, and we find a difference of below 1 Hz to the shifts at a BBR temperature of 300 K given in Table 1 of [14]. At 290(10) K, which we estimate as the temperature of the BBR in the 2S-6P spectroscopy region, we calculate the BBR-induced shift of the 2S-6P$_{1/2}$ and 2S-6P$_{3/2}$ transitions to be $-0.28(1)$ kHz. $\nu_{1/2}$ and $\nu_{3/2}$ have been corrected for the shift by subtracting this value, with the correction assumed to be fully correlated between the transitions. The BBR-induced broadening of the transitions, found to be 1.3 kHz at 300 K in [14], is negligible here.

## 2.3 Zeeman shift

The Zeeman shift of the observed transition frequency for a magnetic flux density $B_x$ along the spectroscopy laser beams is given by (see Section 6.2.4.3 of [1])

$$\Delta\nu_{\text{Zeeman}} = \frac{S_3}{S_0} \frac{g_F \mu_B B_x}{h}. \tag{2}$$

$g_F = 1/3$ ($g_F = 5/3$) is the g-factor of the 6P$_{1/2}^{F=1}$ (6P$_{3/2}^{F=1}$) level, and $\mu_B$ is the Bohr magneton (note that $g_F = 0$ for the 2S$_{1/2}^{F=0}$ level). $S_3/S_0$ is the residual circularly polarized light fraction of the spectroscopy laser beams, where $S_0$ is the total intensity and $S_3$ the intensity difference between right and left circularly polarized light [15, 16].

The magnetic field in the 2S-6P spectroscopy region is minimized with three orthogonal pairs of Helmholtz coils outside the vacuum chamber, and a single-layer, high-permeability metal (mu-metal) shield inside the vacuum chamber. All components inside this shield are made from non-magnetic materials. $B_x$ was measured to be below 1 mG within a 5 mm-radius sphere centered in the spectroscopy region. The absolute value of $S_3/S_0$ in the spectroscopy region was monitored in situ by measuring the polarization of the light backcoupled into the fiber of the active fiber-based retroreflector, as detailed in [15, 16]. For the data set presented here, $|S_3/S_0|$ is 4.2(1.3) % on average (standard deviation over the data groups in parentheses). No polarization data are available for a small number of line scans (approximately 9 %), and we instead use the upper limit of $|S_3/S_0|$ of 10 % from the available data.

Using the value $|S_3/S_0|$ of each line scan (or the upper limit where necessary), and the maximum value of $B_x$ as given above, results in upper limits for the Zeeman shifts of $|\Delta\nu_{\text{Zeeman}}| = 0.02$ kHz and $|\Delta\nu_{\text{Zeeman}}| = 0.11$ kHz for the 2S-6P$_{1/2}$ and 2S-6P$_{3/2}$ transitions, respectively. We include these upper limits as uncertainties for $\nu_{1/2}$ and $\nu_{3/2}$. The uncertainties are assumed to be fully correlated between the 2S-6P$_{1/2}$ and 2S-6P$_{3/2}$ transitions.

## 2.4 Pressure shift

The collision of hydrogen atoms with nearby particles during the 2S→6P excitation leads to a pressure shift of the observed transition frequency. We distinguish between, on the one hand, intra-beam collisions with other hydrogen atoms (either in the 1S or 2S level) or hydrogen molecules in the atomic beam, and, on the other hand, collisions with particles from the background gas. The particles in the atomic beam are assumed to be at the temperature of the nozzle $T_N$, and the background gas is taken to be at room temperature.

We estimate the number of 1S atoms and hydrogen molecules leaving the nozzle per second in the direction of the 2S-6P spectroscopy region to be, respectively, $1.6 \times 10^{16}$ atoms/s and $1.8 \times 10^{16}$ molecules/s, with the latter corresponding to a flux density of $1.4 \times 10^{17}$ molecules/(s/m$^2$) at the spectroscopy laser beams (see Section 4.5.2.3 of [1]). The background gas pressure within the spectroscopy region is estimated (from pressure measurements and conductance simulations) to be $2 \times 10^{-7}$ mbar and is dominated by molecular hydrogen.

A Monte Carlo simulation of the pressure shift expected for the 2S-6P$_{1/2}$ and 2S-6P$_{3/2}$ transitions from intra-beam collisions with other hydrogen atoms, using the approximate parameters and geometry of the 2S-6P measurement, has been done in [17], based on recent calculations of the relevant van der Waals interaction coefficients $C_6$ [18, 19]. Scaling these results to the tenfold lower flux of hydrogen atoms actually used here, we find the magnitude of this contribution to the pressure shift to be below 1 Hz for both transitions and for all velocity groups.

To estimate the pressure shift from molecular hydrogen, we evaluated the collisional shift cross section and then computed the shift using the estimated pressure and velocity distribution of molecules. The velocity-dependent collisional shift cross section was calculated using the S-matrix approach [20], with the matrix elements found by numerical solution of the Schrödinger equation describing evolution of atomic wave function during the collision and subsequently averaged in the collision parameter space by Monte Carlo integration. Finally, the pressure shift for the experimental conditions is calculated from the cross sections with a Monte Carlo simulation. We use a simplified model of the hydrogen molecule which takes into account both the van der Waals interaction with the perturbed hydrogen atom and the quadrupole electric field of the molecule [21]. The evaluation of the van der Waals interaction is complicated by the presence of dipole-allowed transitions in the hydrogen molecule with energy close to the 6P-1S transition in atomic hydrogen. To calculate the non-resonant van der Waals interaction, we use a simplified set of transitions in the hydrogen molecule proposed in [22], giving a van der Waals interaction coefficient of $C_6 = 3.08 \times 10^4$ au (atomic units) for both the 2S-6P$_{1/2}$ and 2S-6P$_{3/2}$ transitions. The close-to-resonance term is found to be negligible due to the small oscillator strength of the corresponding transitions in the hydrogen molecule. In addition, the close-to-resonance energy levels in the molecule may even suppress the pressure shift via transfer of excitation from the hydrogen atom to the molecule during the collision. The resulting pressure shift from intra-beam collisions with hydrogen molecules depends on the atoms' speed but is at most $-1$ Hz for the speeds relevant here ($<1000$ m/s). The resulting pressure shift from the background gas of room-temperature hydrogen molecules, on the other hand, is mostly independent of speed below 1000 m/s, and is at most $-11$ Hz (assuming that the background gas exclusively consists of hydrogen molecules). For both cases, the contribution from the van der Waals interaction dominates over that from the quadrupole electric field of the molecules.

Similarly (and similar to [23]), we estimate the pressure shift from water in the background gas, since the dipole moment of water molecules leads to a comparatively long-range and strong interaction with hydrogen atoms. We find a pressure shift of at most $-3$ Hz for the estimated partial pressure of water ($7 \times 10^{-8}$ mbar) inside the spectroscopy region.

Based on these results, we account for the pressure shift of the 2S-6P$_{1/2}$ and 2S-6P$_{3/2}$ transitions by including a 0.02 kHz uncertainty (assumed to be fully correlated between the two transitions) in $\nu_{1/2}$ and $\nu_{3/2}$.

### 2.5 Sampling bias

A sampling bias, i.e., a dependence of the extracted resonance frequency on the choice of line sampling, can arise when the line shape model used to fit the experimental line shape (and thereby extract the resonance frequency) does not match the experimental line shape exactly. This is the case here, as the Voigt and Voigt doublet line shape models we use for fitting do not include non-Gaussian (or non-Lorentzian) broadening and saturation effects, which are symmetric about the line center, nor do they include asymmetries from the LFS and QI, all of which are present in the experiment. Both symmetric and asymmetric deviations between the line shape fits and the experimental data are clearly visible in the (averaged) fit residuals (see Extended Data Fig. 1), with symmetric deviations dominating over asymmetric ones. Our simulations, on the other hand, include all the above effects. The better agreement of the simulations with the experimental line shapes is visible in Extended Data Fig. 1, which shows similar residuals for line shape fits to the simulations and for line shape fits to the experimental data.

To estimate the sampling bias, we therefore use our simulated line shapes, which are sampled and fit in the same way as the experimental line shapes. We distinguish two cases: a sampling bias for a line shape with asymmetric deviations sampled symmetrically about its unperturbed line center (i.e., the line center in the absence of asymmetric deviations), and a sampling bias for a line shape with generally both asymmetric and symmetric deviations sampled not symmetrically but offset from its unperturbed line center. Since we define the corrections for the LFS and the QI shift as the resonance frequency from a fit to the corresponding simulations, the first case is intrinsically included in those corrections. For the second case, we apply a frequency offset to the sampling of the simulations. The resulting bias is linear in, but of opposite sign to, the frequency offset (for offsets $\lesssim 20\,\text{kHz}$), and identical within $10\,\text{Hz}$ for the LFS simulations, QI simulations, and simulations using a simple model that includes non-Gaussian broadening and saturation, but not LFS or QI asymmetries. Hence, the bias is dominated by symmetric deviations, and we use the simple model to estimate it. We find the average frequency offset for each data group from the difference between the center laser frequency and the transition frequencies determined in this work. The average frequency offsets are mostly positive (caused by drift of the spectroscopy laser frequency) and reach at most $15\,\text{kHz}$, while the resulting bias reaches at most $-0.19\,\text{kHz}$. Overall, we estimate the sampling bias to be $-0.04\,\text{kHz}$ and $-0.08\,\text{kHz}$ for the 2S-6P$_{1/2}$ and 2S-6P$_{3/2}$ transitions, respectively. We include these estimates as uncertainties for $\nu_{1/2}$ and $\nu_{3/2}$ and treat them as fully correlated between the two transitions.

## 2.6 Signal background

We observe a signal background $y_0$, i.e., a signal when the 2S-6P spectroscopy laser is off resonance, on both detectors. The relative signal background $y_0/A$, where $A$ is the line amplitude above the background when the spectroscopy laser is on resonance, ranges from $3\,\%$ to $12\,\%$ for the different data groups, and scales approximately inversely with spectroscopy laser power. It is approximately constant for the different velocity groups, the same for both detectors, and consistent over freezing cycles (and therefore atomic beam offset angle alignments). We attribute the signal background to Lyman-$\alpha$ photons emitted by the decay of metastable 2S atoms, as it is only present when atomic hydrogen flows into the nozzle and the 1S-2S preparation laser is on resonance (the dark count rate of the detectors when no 2S atoms are present is negligible).

The spontaneous decay of 2S atoms to the 1S ground level (by two-photon emission with a rate of $8.2\,\text{s}^{-1}$) inside the 52-mm-diameter 2S-6P spectroscopy region does not contribute substantially to the observed signal background. This is because, while the ratio of decayed 2S atoms to 2S atoms that have been excited to the 6P level can reach up to $50\,\%$ of the observed value of $y_0/A$ (see Tables 5.1 and 5.4 of [1]), the detection efficiency of the emitted 243-nm photons is approximately four orders-of-magnitude lower than for the signal (Lyman-$\epsilon$) photons.

External electric fields can lead to one-photon (Lyman-$\alpha$) decay, or quenching, of 2S atoms through mixing of the 2S and 2P levels (with Lyman-$\alpha$ photons having a five-fold lower detection efficiency than Lyman-$\epsilon$ photons). The observed stray electric fields (strength below $1\,\text{V/m}$) inside the spectroscopy region, however, only lead to a one-photon decay rate of $3\times 10^{-3}\,\text{s}^{-1}$, much too low to explain the signal background. Larger stray fields might be present at the apertures of the Faraday cage where the atomic beam enters and exits the spectroscopy region, but we have not characterized the stray fields outside the spectroscopy region.

Collisions with nearby particles can likewise quench the 2S atoms or deflect them (by angles $>0.1\,\text{rad}$) towards surfaces inside the spectroscopy region, where they quench upon impact. The collision partners and relevant numbers are discussed for the closely-related pressure shift in Section 2.4. We estimate that intra-beam collisions with 1S atoms and molecular hydrogen can only explain $0.4\,\%$ and $2\,\%$, respectively, of the signal background. We find the dominant process to be collisional quenching by the quadrupole electric field of the hydrogen molecules [24] in the room-temperature background gas. However, the background gas density inside the spectroscopy region only explains $\approx 10\,\%$ of the observed signal background. We hypothesize that the bulk of the signal background stems from 2S atoms rapidly being quenched by molecular hydrogen as they leave the spectroscopy region and enter the outer vacuum region through a 30-mm-long differential pumping tube (see Section 4.2 of [1]), as the background gas pressure in the outer vacuum region is 30-fold higher than in the spectroscopy region but likewise dominated by molecular hydrogen.

This hypothesis is supported by the finding that the relative signal background scales linearly with the background gas pressure. As the fraction of molecular hydrogen freezing inside the nozzle depends strongly on temperature, we observed this scaling by varying the nozzle temperature within $3.7\,\text{K}\ldots 6\,\text{K}$. This changed the background gas pressure and the relative signal background by more than an order of magnitude (see Section 4.5.2 of [1]). Furthermore, as the background pressure of water is not expected to strongly vary with the nozzle temperature, this test excludes quenching by water molecules as a dominant source for the signal background. For the data set presented here, for which the nozzle was held at constant temperature and the background gas pressure varied by less than 15 % (relative standard deviation), we find a significant correlation of the relative signal background with the pressure ($r = 0.50$ and $r = 0.29$ for the spectroscopy region and outer vacuum region, respectively). There is no significant correlation with the Doppler-free transition frequency $\nu_e$.

The hypothesis is also consistent with the relative signal background being constant over velocity groups, as the speed of the background gas particles is always much larger than that of the 2S atoms, leading to an approximately constant relative speed. In this case, the collision probability scales with the time spent in the background gas, and therefore with $1/v$, where $v$ is the atom's speed. Since the 6P excitation probability also approximately scales as $1/v$ for our experimental conditions, this results in a relative signal background independent of $v$, i.e., the relative signal background is identical for the different velocity groups.

A signal background that is not constant with spectroscopy laser detuning could potentially lead to line shifts. To constrain such shifts, we estimate the linear slope of the signal background from the difference in the signal at the maximum detunings ($\Delta = \pm 50$ MHz), where we take into account any signal differences expected from the LFS or QI shift. We then artificially remove this linear slope from the signal of each line scan and re-analyze the data. Overall, we find statistically insignificant shifts of, respectively, $-40$ Hz and $-30$ Hz for the 2S-6P$_{1/2}$ and 2S-6P$_{3/2}$ transitions. We include these estimates as uncorrelated uncertainties for $\nu_{1/2}$ and $\nu_{3/2}$.

## 2.7 Laser systems and frequency standard

The two laser systems used in the measurement, the 2S-6P spectroscopy laser at 410 nm and the 1S-2S preparation laser at 243 nm, share a similar design. Both systems use external cavity diode lasers (cavity length ∼20 cm, described in [25]) as seed lasers in the infrared at 820 nm and 972 nm. The lasers are phase-stabilized to high-finesse Fabry-Pérot cavities, which reduces the linewidth to a few Hz [26]. This narrow carrier sits on a weak but broad (∼1 MHz) noise pedestal [25, 26]. The linewidth and the spectral purity of the lasers are monitored with a beat note between them using a low-noise, Er-doped fiber optical frequency comb as transfer oscillator [27, 28]. After power amplification with a tapered amplifier, the light is frequency doubled (frequency doubled twice) to 410 nm (243 nm) for the spectroscopy (preparation) laser.

Asymmetries in the noise pedestal of the spectroscopy laser may lead to line shifts. We obtain an upper limit of 0.07 kHz by artificially introducing an asymmetry of 10 % into the measured noise pedestal [25], numerically convolving it with the 2S-6P line shape, and fitting the result. We include this upper limit as a fully correlated uncertainty for the 2S-6P$_{1/2}$ and 2S-6P$_{3/2}$ transitions.

The frequency comb is used to compare the optical frequencies of the 2S-6P spectroscopy laser and the 1S-2S preparation laser to the microwave frequency of a passive hydrogen maser (see Section 4.8 of [1] and Section 3.2.3 of [29]). To this end, the repetition rate and carrier–envelope offset frequency of the frequency comb, and the beat notes between the frequency comb modes and the lasers, are continuously measured using frequency counters referenced to the maser. The frequency of the hydrogen maser, in turn, is continuously compared against the caesium frequency standard, which is the basis of the unit of hertz in the International System of Units (SI), using a global navigation satellite system (GNSS) receiver (with the gravitational redshift taken into account). This comparison leads to corrections of the transition frequencies by 0.02 kHz ($2.5 \times 10^{-14}$) and 0.01 kHz ($1.9 \times 10^{-14}$) for the 2S-6P$_{1/2}$ and 2S-6P$_{3/2}$ transitions, respectively. The total fractional frequency uncertainty of the comparison (between the frequency comb, hydrogen maser, and caesium standard) is estimated as $1 \times 10^{-14}$, corresponding to 7 Hz for both transitions (here rounded to 0.01 kHz and assumed to be fully correlated).

The fundamental frequencies (in the infrared) of the spectroscopy and preparation lasers drift, on average, by 5.4(2.9) kHz/day and 2.6(2.7) kHz/day, respectively (standard deviation over measurement days in parentheses). During the approximately 1 min required to record a single line scan, this corresponds to an insignificant drift at the atomic transition frequencies of 8 Hz and 15 Hz. We determine the laser frequencies by linear fits to 1-hour-long windows of the frequency counter data, resulting in a statistical uncertainty below 10 Hz for each window, and therefore negligible compared to the statistical uncertainty of the line shape fits.

## 2.8 Recoil shift

Energy and momentum conservation require the energy of a photon driving an atomic transition of frequency $\nu$ to be larger than the corresponding transition energy $h\nu$ by the recoil shift. The recoil shift for the for 2S-6P$_{1/2}$ and 2S-6P$_{3/2}$ transitions can be written as

$$\Delta\nu_{\rm rec} = \frac{h}{2m_{\rm H}} \left(\frac{\nu}{c}\right)^2 \approx 1176.03\,{\rm kHz}, \tag{3}$$

The transition frequency $\nu$ is identical at the required level of accuracy for both transitions and is known with much lower uncertainty than needed here. Likewise, the mass of the hydrogen atom can be found by $m_{\rm H} \approx m_{\rm e}(m_{\rm p}/m_{\rm e} + 1 + \alpha^2/2)$, where $m_{\rm e}$ is the electron mass, with sufficient accuracy. Corrections to $\Delta\nu_{\rm rec}$ from wavefront curvature and the Gouy phase [30] amount to at most 0.1 Hz and are neglected here. $\nu_{1/2}$ and $\nu_{3/2}$ have been corrected for the recoil shift by subtracting $\Delta\nu_{\rm rec}$.